\documentclass[11pt,a4paper]{article}
\pdfoutput=1
\usepackage{paperstyle}

\usepackage{amsmath,amssymb,graphicx}
\usepackage{soul}

\usepackage{color}

\newcommand{\be}{\begin{eqnarray}}
\newcommand{\en}{\end{eqnarray}}

\newcommand{\fnl}{f_{\rm{NL}}}
\newcommand{\mpl}{m_{\rm{pl}}}
\newcommand{\ns}{n_{\rm{s}}}
\newcommand{\As}{A_{\rm{s}}}
\newcommand{\Ne}{N_{\rm{e}}}
\newcommand{\Hubble}{H}
\newcommand{\Del}{\Delta}
\newcommand{\braket}[1]{\langle #1 \rangle}
\newcommand{\kahler}{K\"{a}hler\ }

\newcommand{\phiI}{\overline{\phi}}
\newcommand{\chiI}{\overline{\chi}}

\newcommand{\alphaBetaThr}{1.4}

\title{Hilltop inflation with preinflation from coupling to matter fields }
\author[a,b]{Stefan Antusch,}
\author[a]{David Nolde}
\author[a]{and Stefano Orani}
\affiliation[a]{
Department of Physics, University of Basel,\\
Klingelbergstrasse\ 82, CH-4056 Basel, Switzerland}
\affiliation[b]{Max-Planck-Institut f\"ur Physik (Werner-Heisenberg-Institut),\\
F\"ohringer Ring 6, D-80805 M\"unchen, Germany}
\emailAdd{stefan.antusch@unibas.ch}
\emailAdd{david.nolde@unibas.ch}
\emailAdd{stefano.orani@unibas.ch}
\abstract{
We propose a class of models of supersymmetric hilltop inflation (also called ``new inflation'') where the initial conditions of the inflaton close to the hilltop are generated through ``matter field preinflation''. This is achieved via a coupling term between the inflaton and matter fields (i.e. Standard Model fields or a right-handed neutrino). The same coupling also opens up a decay channel for the inflaton into Standard Model fields, which allows efficient reheating of the universe. We discuss the multifield dynamics of the inflaton and matter fields during inflation using the $\delta N$ formalism and show under which conditions the model effectively reduces to single-field hilltop inflation during the last 60 e-folds. We also study perturbative reheating through the matter-inflaton coupling for a specific example where the matter field is identified with a right-handed (s)neutrino, and demonstrate that in this case the model can generate the observed baryon asymmetry through nonthermal leptogenesis.
}

\begin{document}

\maketitle

\section{Introduction}

Inflation provides a successful paradigm for the early universe, capable not only of resolving the flatness and horizon problems of big bang cosmology, but also of generating primordial perturbations which eventually develop into the present day inhomogeneities of the universe and can be probed by observations of the CMB and large scale structures.

The Planck satellite has recently observed the temperature anisotropies of the CMB with unprecedented precision~\cite{Ade:2013zuv,Ade:2013uln,Ade:2013ydc}. The new data has brought further confirmation of inflation and provided useful information towards discriminating between different models of inflation. In particular, the non-Gaussianity parameter has been constrained to $f_{\text{NL}} = 2.7 \pm 5.8$ at $68\%$ C.L., which is consistent with single-field slow-roll inflation. The spectral index $\ns$ has been measured to be $\ns = 0.9603\pm0.0073$ at $68\%$ C.L., which constrains the possible shapes of the inflaton potential, and the upper bound on the tensor-to-scalar ratio has been tightened to $r < 0.11$ at $95\%$ C.L., which already excludes some large-field models of inflation.

Among the various inflation scenarios, hilltop inflation \cite{hilltop,newinf,newinfSG,Senoguz:2004ky} (also called ``new inflation'') can lead to a primordial spectrum close to Planck's best-fit value \cite{Ade:2013uln}. In hilltop inflation, the inflaton field rolls down from a maximum of a very flat potential which is shaped like a plateau. Such potentials can arise when a phase transition takes place, for example associated with the breaking of a GUT symmetry \cite{newinf} or a flavour symmetry \cite{Antusch:2008gw}, which allows for possible close connections to particle physics.

However, given that the inflaton field $\phi$ has to start its dynamics close to the maximum of its potential, the obvious question arises of how the field initially gets there. Many of the scenarios proposed to resolve this initial condition problem introduce a preceding period of slow-roll inflation (called ``preinflation'') driven by a preinflaton $\chi$. During this phase of preinflation, the inflaton field $\phi$ is forced to the maximum of its potential, which dynamically generates the initial conditions required for hilltop inflation in $\phi$. Usually, this involves adding a new sector of singlet fields which are decoupled from the visible sector (e.g. in \cite{Izawa:1997df,Senoguz:2004ky,smoothPreinflation}). A similar function is performed by a coupling of the inflaton to the energy density as proposed in \cite{symmetron}. Other approaches to the initial conditions problem include \cite{otherICs}.

In this paper, we show how preinflation can be realized in supersymmetric models using only a coupling of the inflaton to ``matter fields'', which we define as Standard Model fields or right-handed neutrinos (and their superpartners). In the stage of preinflation, the scalar component of a matter superfield (or alternatively of a D-flat combination of charged matter superfields) takes the role of the preinflaton and generates a mass term for the inflaton which leads to the right initial conditions for hilltop inflation. After hilltop inflation, the same coupling to matter fields allows the inflaton to decay and reheat the universe.

We also study the effects of the preinflaton matter field on the primordial spectrum, and we illustrate how reheating can occur via the inflaton-matter coupling for the example of a right-handed sneutrino preinflaton.

The paper is organized as follows: in sections~\ref{sec:idea} and~\ref{sec:model} we discuss qualitatively the main ideas of the paper and present the model. We then calculate in section~\ref{sec:ics} the likely initial conditions that arise from quantum fluctuations near the instability point. Using these initial conditions, in section~\ref{sec:numericalResults} we numerically calculate the spectrum of perturbations from inflation using the $\delta N$ formalism (briefly introduced in section~\ref{sec:deltaN}), confirming that single-field hilltop inflation preceded by matter field preinflation is a generic outcome of our model setup for a large region of parameter space. Finally, in section~\ref{sec:reheating} we demonstrate for a specific example model that the inflaton-matter coupling can provide a viable channel for perturbative reheating and leptogenesis.

\section{Hilltop Inflation: Challenges and Main Idea of the Paper}
\label{sec:idea}

Models of hilltop inflation are usually presented in form of a potential for a real scalar field $\phi$,
\be
V(\phi) = \Lambda^4 \left(1 - 2\frac{\phi^p}{\mu^p} + \dots \right) \:.
\en
The dots contain terms with higher power of $\phi$. With $p\ge 4$, the model is in very good agreement with the recent CMB observations by Planck \cite{Ade:2013uln}. Inflation is realised when the inflaton field $\phi$ rolls slowly from close to the top of the potential with $\phi\approx 0$, where the potential has the shape of a very flat plateau, towards its global minimum at $\braket{\phi} < \mpl$. However, in order to construct a convincing model of hilltop inflation, the following challenges have to be resolved:

\begin{enumerate}
\item With $\phi$ being a real scalar field, how can one suppress terms of the form $\Lambda^4 \frac{\phi^q}{\mu^q}$ with $q < 4$ in $V(\phi)$? Such terms (unless they have coefficients tuned to be very small) would lead to predictions inconsistent with observations.
\item For obtaining enough e-foldings of inflation, the field $\phi$  has to start quite close to the top of the potential. But how does the field initially get there?
\end{enumerate} 

{\bf Challenge 1} can be overcome when the model is constructed in a supersymmetric (i.e.\ supergravity) framework. With fields being promoted to superfields, the simple superpotential 
\be
W_{\rm hilltop} = \hat{S} \left(\frac{\hat{\Phi}^p}{M^{p - 2}} - \Lambda^2\right)
\en
generates (with $S=0$ due to supergravity corrections, and taking $M$ and $\Lambda$ real without loss of generality) the following scalar potential for the scalar component $\Phi$ of the superfield $\hat{\Phi}$:
\be
V(\Phi) = | F_S |^2 = \left| \frac{\partial W}{\partial \hat{S}} \right|^2_{\theta = 0} = \left|\frac{\Phi^p}{M^{p - 2}} - \Lambda^2\right|^2 
= \Lambda^4 \left(1 - 2\frac{\mathrm{Re}(\Phi^p)}{\tilde{\mu}^p} + \frac{|\Phi |^{2p}}{\tilde{\mu}^{2p}} \right)
\en 
with $\tilde{\mu}^p :=  M^{p - 2} \Lambda^2$ and where $\Phi$ is now a complex scalar field which may be decomposed into real and imaginary parts as $\Phi = \frac{1}{\sqrt{2}}( \phi + i \phiI )$. This has the desired form, with additional terms containing only higher powers of $\phi$. 

The advantage of the supersymmetric formulation is that one can now easily fix this form of the superpotential by demanding symmetries, i.e.\ a $U(1)_\mathrm{R}$ symmetry and a $\mathbb{Z}_p$ symmetry, when we distribute two units of $U(1)_\mathrm{R}$ charge to $\hat{S}$ and zero to $\hat{\Phi}$, and one unit of $\mathbb{Z}_p$ charge to $\hat{\Phi}$ while $\hat{S}$ is a $\mathbb{Z}_p$-singlet.\footnote{Also, supergravity corrections from the K\"ahler potential can be further suppressed (in addition to the Planck scale suppression) by e.g.\ a Heisenberg symmetry, as discussed in \cite{Antusch:2013eca}.} Furthermore, this can also be generalized to the case where $\hat\Phi$ is charged under a general (global or local) symmetry group $G$.\\

{\bf Challenge 2} can be resolved by an epoch of so-called ``preinflation'' \cite{Izawa:1997df,Senoguz:2004ky,smoothPreinflation}, where the dynamics of other fields force $\phi$ to be close to zero (i.e.\ to the hilltop). This generally involves some couplings between $\phi$ and other scalar fields $\chi_i$ which we may summarise as $f(\phi,\chi_i)$ such that the potential reads
\be
V(\phi) = \Lambda^4 \left(1 - 2\frac{\phi^p}{\mu^p} + \dots \right) +  f(\phi,\chi_i) \:.
\en 
$f(\phi,\chi)$ has to be constructed such that  $\phi \approx 0$ before the observable part of inflation starts.

The {\bf main idea of this paper} is to extend the superpotential $W_{\rm hilltop}$ by a simple term which couples the inflaton $\phi$ to a matter superfield $\hat{X}$ (containing a Standard Model particle and its superpartner), namely
\be
W = \hat{S} \left(\frac{\hat{\Phi}^p}{M^{p - 2}} - \Lambda^2\right) + \lambda\hat{\Phi}^2 \hat{X}^n.
\label{wgen}
\en 
$\hat{X}^n$ can be any gauge invariant contraction of matter superfields, e.g.\ the right-handed neutrino superfield which contains the right-handed neutrino fermion and the scalar right-handed sneutrino, or some lepton-Higgs ($\hat{L} \hat{H}_u$, $\hat{L}\hat{H}_d\hat{E}$) or quark-Higgs ($\hat{Q} \hat{H}_u \hat{U}$, $\hat{Q} \hat{H}_d \hat{D}$) direction.\footnote{Throughout this paper, we treat $\hat{X}$ as a single field. Replacing it with a contraction of fields $\hat{X}^2 \rightarrow \hat{X}_1 \hat{X}_2$ does not change our results except for combinatorial factors of $O(1)$.}

Such a coupling has the following advantages, in addition to being able to fix the shape of the superpotential by symmetries (i.e.\ to {\bf solve challenge 1} in a supersymmetric framework):

\begin{itemize}
\item With the scalar component $X$ of $\hat{X}$ being non-zero in an early epoch, a mass term for $\phi$ is generated which drives it to small field values. This yields a stage of {\bf preinflation} (along the lines of so-called ``tribrid inflation'' models \cite{sneutrinoTribrid,gnsTribrid,kahlerTribrid}) and thus {\bf solves challenge 2} by a rather minimalistic extension of the model, using only matter fields for preinflation, where by matter fields we mean Standard Model fields and right-handed neutrinos as well as their superpartners.

\item After inflation the inflaton $\phi$ decays by the same coupling to the components of the matter superfield $\hat{X}$, which decay further into the other fields of the Standard Model (or MSSM) and efficiently {\bf reheat the universe}. When $\hat{X}$ is a right-handed neutrino superfield, the latter decay would proceed via the neutrino Yukawa couplings, and can even generate the baryon asymmetry of the universe via the non-thermal leptogenesis mechanism \cite{leptogenesis}.
\end{itemize}

\section{The Model}
\label{sec:model}

In what follows we focus on the choices $p=4$ and $n=2$ in the superpotential \eqref{wgen}. Larger values of $p$ increase the flatness of the plateau, while larger $n$ leads to higher order interactions between the scalar fields $\Phi$ and $X$.

Our choices lead to the scalar potential, with $S=0$ due to supergravity corrections (see appendix~\ref{appendix:scalarPotential} for details):
\begin{align}
 V &= \left| \frac{\partial W}{\partial \hat{S}} \right|^2_{\theta = 0} + \left| \frac{\partial W}{\partial \hat{\Phi}} \right|^2_{\theta = 0} + \left| \frac{\partial W}{\partial \hat{X}} \right|^2_{\theta = 0} + V_\text{SUGRA} \notag\\
 &= \Lambda^4\left(\frac{\phi^4}{\mu^4}-1\right)^2+\frac{1}{2}\lambda^2\phi^2
\chi^2(\phi^2+\chi^2)+\frac{1}{2}m^2_\chi\chi^2-\frac{1}{2}m^2_\phi\phi^2\,,
\label{vinf}
\end{align}
where $\mu^2=2M\Lambda$ and $m^2_\chi>0$, $m^2_\phi>0$ are masses induced by supergravity corrections. The fields are $\phi=\sqrt{2} \operatorname{Re} [\Phi]$ and $\chi=\sqrt{2} \lvert X \rvert$. The imaginary part of $\Phi$ can affect the dynamics and observational signatures of hilltop inflation, as discussed in \cite{Nolde:2013bha}. We first restrict our discussion to $\operatorname{Im} [\Phi] = 0$ and discuss the effect of the imaginary part later in section \ref{sec:imaginaryInflatonEffects}.

It will be useful to parametrize the masses $m^2_\chi$ and $m^2_\phi$ as
\begin{align}
 m^2_\chi = \alpha \Lambda^4/\mpl^2, \quad m^2_\phi = \beta \Lambda^4/\mpl^2.
\end{align}
In order to have $\phi$ as the inflaton and $\chi$ as the preinflaton, both fields must be able to slow-roll and therefore have masses below the Hubble scale. With $\Hubble \simeq \Lambda^2/(\sqrt{3}\mpl)$, this requires $\alpha$, $\beta < 1/3$. For $\beta$, we will later find the stronger constraint $\beta \lesssim 0.03$ to produce the correct spectral index $\ns$ as measured by the Planck satellite.

The mass of $\phi$ at $\phi=0$ is given by $\lambda^2\chi^4-m^2_\phi$, and it becomes tachyonic when $\chi<\chi_{\rm crit}\equiv\sqrt{m_\phi/\lambda}$, defining the critical value $\chi_{\rm crit}$. When $\chi<\chi_{\rm crit}$, $\phi=0$ becomes unstable and the field trajectory rolls to one of the two global minima of \eqref{vinf}, at $\chi=0$ and $\phi=\pm \mu$. This symmetry breaking can generate topological defects, which can affect the evolution of spacetime. However, in our scenario, the symmetry breaking is followed by a long phase of hilltop inflation, therefore stretching such relics outside of observable scales (for this to happen, accelerated expansion after the symmetry breaking must last at least $60$ e-folds).

The inflationary dynamics are as follows: initially, $\mpl>\chi>\chi_{\rm crit}$ and $\phi=0$. The potential is of the form 
\be 
V_{\rm pre} \simeq \Lambda^4 + \frac{1}{2}m^2_\chi\chi^2\,,
\en
with $2\Lambda^4\gg m^2_\chi\chi^2$. $\chi$ rolls towards $\chi_{\rm crit}$ and many e-folds of accelerated expansion are generated during this preinflation phase.

As $\chi>\chi_{\rm crit}$ approaches the instability, $\phi$'s mass decreases and eventually it becomes light enough for the amplitude of its fluctuations about $\phi=0$ to start growing. For values of $m^2_\chi$ which are not excessively small, the motion along the $\chi$ direction while going through the instability is dominated by the field's classical rolling. On the other hand, $\phi$'s fluctuations are quantum inside a diffusion region defined by 
\be 
\left|\frac{\dot{\phi}}{H}\right|^2<\left(\frac{H}{2\pi}\right)^2\,.
\en
The field $\phi$ exits the diffusion region at some $\chi<\chi_{\rm crit}$ when its classical motion takes over and it rolls away from the origin. Then a second phase of classical inflation starts, driven by the potential \eqref{vinf}. This second phase is initially dominated by the lowest order terms in $\phi$. As $\phi$ grows, eventually the $\phi^4$ term starts dominating and inflation ends when the oscillating phase about one of the two global minima $\chi=0$, $\phi=\pm\mu$ starts. Since more than $60$ e-folds of accelerated expansion are realised in the second phase, observables exclusively depend on its dynamics and the first phase only provides a mechanism for generating the initial conditions for the hilltop inflation phase.

After inflation, the universe reheats. In our model, the inflaton $\Phi$ decays to the scalar and fermionic components of the matter superfield $\hat{X}$ via its superpotential coupling. These can then decay further into the Standard Model particles and their superpartners via gauge interactions or Yukawa couplings, depending on which matter superfield is used for $\hat{X}$.

\section{On the initial conditions near the instability}
\label{sec:ics}

\begin{figure}[bt]
  \centering$
\begin{array}{cc}
\includegraphics[width=0.48\textwidth]{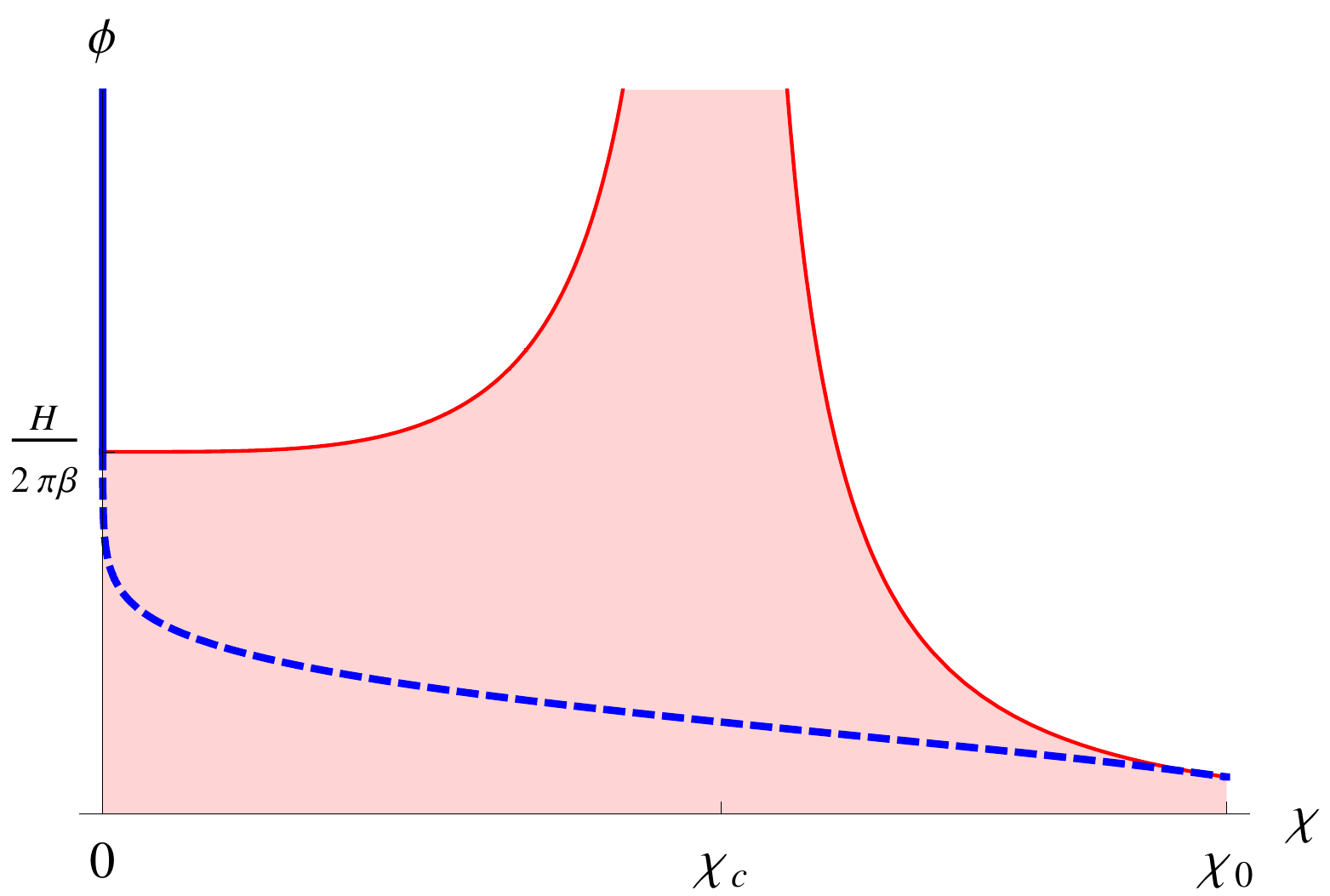} &
\includegraphics[width=0.48\textwidth]{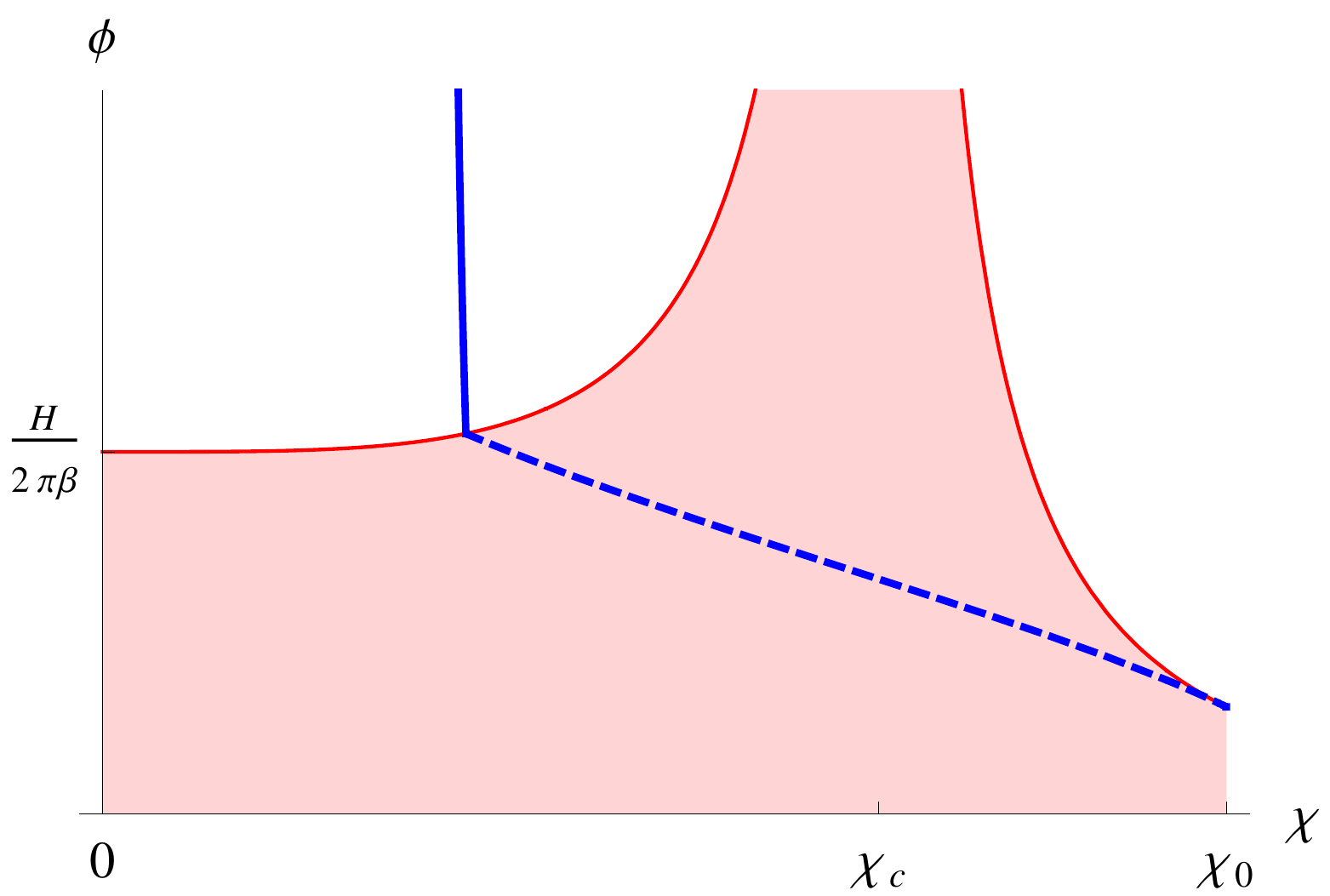}
\end{array}$
  \caption{Field trajectory (blue curve) for $\beta = 0.03$ through the diffusion region (red area) for $\alpha = 10^{-2}$ (left) and $\alpha = 10^{-3}$ (right). The red curve is the diffusion boundary $\phi_\text{b}(\chi)$ given in eq.~\eqref{eq:phiB}. The field trajectory enters the diffusion region at $(\phi_0,\chi_0)$ and leaves it at $(\phi_1,\chi_1)$. The aim of this section is to calculate $(\phi_1,\chi_1)$ which we identify as the initial conditions for the observable part of inflation.
	The left plot is the generic result for $\beta^2 \ll \alpha$, in which case $\chi_1 \rightarrow 0$ and inflation proceeds as single-field inflation in $\phi$. Only for $\beta^2 \gtrsim \alpha$, $\chi_1$ can be significantly large, in which case inflation should be analysed in a multi-field framework like the $\delta N$ formalism.}
  \label{fig:diffusionRegion}
\end{figure}

During preinflation with $\chi \gg \chi_\text{crit}$, the inflaton field $\phi$ is driven to zero by the large mass term $(\frac{\lambda^2}{2} \chi^4 - m_\phi^2) \phi^2$. However, when $\chi$ approaches the instability at $\chi = \chi_\text{crit}$, the inflaton field $\phi$ becomes massless and $\braket{\phi^2}$ grows due to quantum fluctuations. Quantum diffusion dominates when the classical displacement $\delta \phi_{\text{cl}}$ of the field per Hubble time is smaller than the growth of quantum fluctuations $\delta \phi_{\text{qu}}$ per Hubble time $\Hubble^{-1}$:
\begin{align}
 &\delta \phi_{\text{qu}} = \frac{\Hubble}{2\pi} \, \stackrel{!}{>} \, \delta \phi_{\text{cl}}
 \, = \, H^{-1} \lvert \dot{\phi} \rvert \, \simeq \, \left| \frac{\partial_\phi V}{V} \right| \mpl^2.
\end{align}
In this section, we want to discuss how the fields evolve during this phase to find expressions for the field values $\phi_1$ and $\chi_1$ at the end of the quantum diffusion regime. To do this, we take the following steps (see fig.~\ref{fig:diffusionRegion}):
\begin{enumerate}
 \item We find the boundary $\phi_\text{b}(\chi)$ of the diffusion region inside which quantum diffusion dominates over the classical slow-roll field evolution for $\phi$.
 \item Inside this boundary, we assume that $\phi$ grows due to quantum fluctuations. We estimate the value of $\phi$ in this region via the expectation value $\phi_\text{diff} \sim \sqrt{ \braket{\phi^2} }$.\footnote{Of course, quantum fluctuations are random, so the realized value of $\phi$ can be different for different patches of the universe. We use the expectation value only for an estimate of the likely magnitude of $\phi$. This is sufficient because our conclusions will only depend on the order of magnitude of $\phi$. The precise initial conditions are less important as the trajectories exhibit an attractor behaviour towards single-field hilltop inflation.} Outside the boundary, we calculate the trajectory of $\phi$ from the slow-roll equations of motion.
 \item Assuming that initially we start from $\chi \gg \chi_\text{crit}$ (which quickly drives $\phi \rightarrow 0$), we find the values $(\phi_0,\chi_0)$ at which $\phi$ enters the diffusion region and the values $(\phi_1,\chi_1)$ at which it leaves the diffusion region. $\phi_1$ and $\chi_1$ are then the initial condition for inflation using the classical slow-roll equations of motion.
\end{enumerate}

\subsection{Approximate potential during quantum diffusion phase}

To keep the calculation of the diffusion region simple and instructive, we only keep terms of lowest order in $\phi$, i.e.
\begin{subequations}
\begin{align}
 \mpl^2 \partial_\phi V \, &\simeq \, \left( -\beta \Lambda^4 +  \mpl^2 \lambda^2 \chi^4  \right)\phi  \, = \,  -\beta \Lambda^4\left(  1 - \frac{\chi^4}{\chi_\text{crit}^4}  \right)\phi , \\
 \mpl^2 \partial_\chi V \, &\simeq \, \alpha \Lambda^4 \, \chi.
\end{align}
\end{subequations}
This approximation usually works well because $\phi$ is very small in this region, as it is generated from quantum fluctuations only.

One must include the higher-order terms in the calculation only if $\alpha$ or $\beta$ are very small. In that case, the following calculation can be performed numerically using the exact scalar potential. The exact values for $\alpha$ and $\beta$ for which this is necessary depend on $\mu$ and $\lambda$, but they are typically much smaller than about $10^{-4}$. We have checked explicitly that the approximation is very good for the range of parameters that we use in our numerical calculations.

\subsection{Boundary $\phi_\text{b}$ of the diffusion region}

We find a formula for the boundary between the diffusion region and the classical region by inserting the scalar potential $\partial_\phi V$:
\begin{align}
 \frac{\Hubble}{2\pi} \, \stackrel{!}{=} \, \mpl^2 \left| \frac{\partial_\phi V}{V} \right|_{\phi_\text{b}} \, = \, \beta \left| 1 - \frac{\chi^4}{\chi_\text{crit}^4} \right| \phi_\text{b}.
\end{align}
In the following, it will be useful to substitute the field $\chi$ with its quartic displacement from $\chi_\text{crit}$:
\begin{align}
 \Del(\chi) \, = \, \frac{\chi^4}{\chi_\text{crit}^4} - 1. \label{eq:defDel}
\end{align}
Using this variable, the diffusion boundary can be written as
\begin{align}
 \phi_\text{b}(\Del) \, = \, \left( \frac{\Hubble}{2\pi} \right) \frac{1}{\beta \lvert \Del \rvert}. \label{eq:phiB}
\end{align}

\subsection{Trajectory $\phi_\text{diff}$ inside the diffusion region}

After the field enters the diffusion region at some $\Del_0 > 0$, the expectation value for the squared field grows linearly with $N = \Hubble t$. Inside the diffusion region, $\braket{\phi^2}$ is then given by
\begin{align}
 \phi_{\text{diff}}^2( \Del ) \, = \, \phi_\text{b}^2(\Del_0) + \left( \frac{\Hubble}{2\pi} \right)^2 N(\Del). \label{eq:phiDiff1}
\end{align}
The function $N(\Del)$ can be inferred from the slow-roll equation of motion for $\chi$, which is dominated by the term proportional to $m_\chi$:
\begin{align}
 0 \, &\simeq \, 3 H^2 \chi'(N) + \partial_\chi V \, \simeq \, \Lambda^4 \left[ \chi'(N) + \alpha \chi(N) \right].
\end{align}
This differential equation can be solved for $\chi(N)$:
\begin{align}
 \chi(N) \, = \, \chi_0 \, e^{-\alpha N},
\end{align}
which implies, using eq.~\eqref{eq:defDel}:
\begin{align}
 N(\Del) \, = \, \frac{1}{\alpha} \ln \left( \frac{ \chi_0 }{ \chi } \right)
 \, = \, \frac{1}{4\alpha} \ln \left( \frac{ \Del_0 + 1 }{ \Del + 1 } \right). \label{eq:NDel}
\end{align}
Combining eqs.~\eqref{eq:phiB}, \eqref{eq:phiDiff1} and \eqref{eq:NDel} we find the expression for the diffusion boundary in terms of $\Del$:
\begin{align}
 \phi_{\text{diff}}^2( \Del ) \, = \, \left( \frac{\Hubble}{2\pi} \right)^2 \left[  \frac{1}{\beta^2 \Del_0^2}  +  \frac{1}{4\alpha} \ln \left( \frac{ \Del_0 + 1 }{ \Del + 1 } \right)  \right]. \label{eq:phiDiff2}
\end{align}

\subsection{Diffusion region entry at $\Delta_0$ and exit at $\Delta_1$}

\begin{figure}[bt]
\centering
$
\begin{array}{cc}
\includegraphics[width=0.48\textwidth]{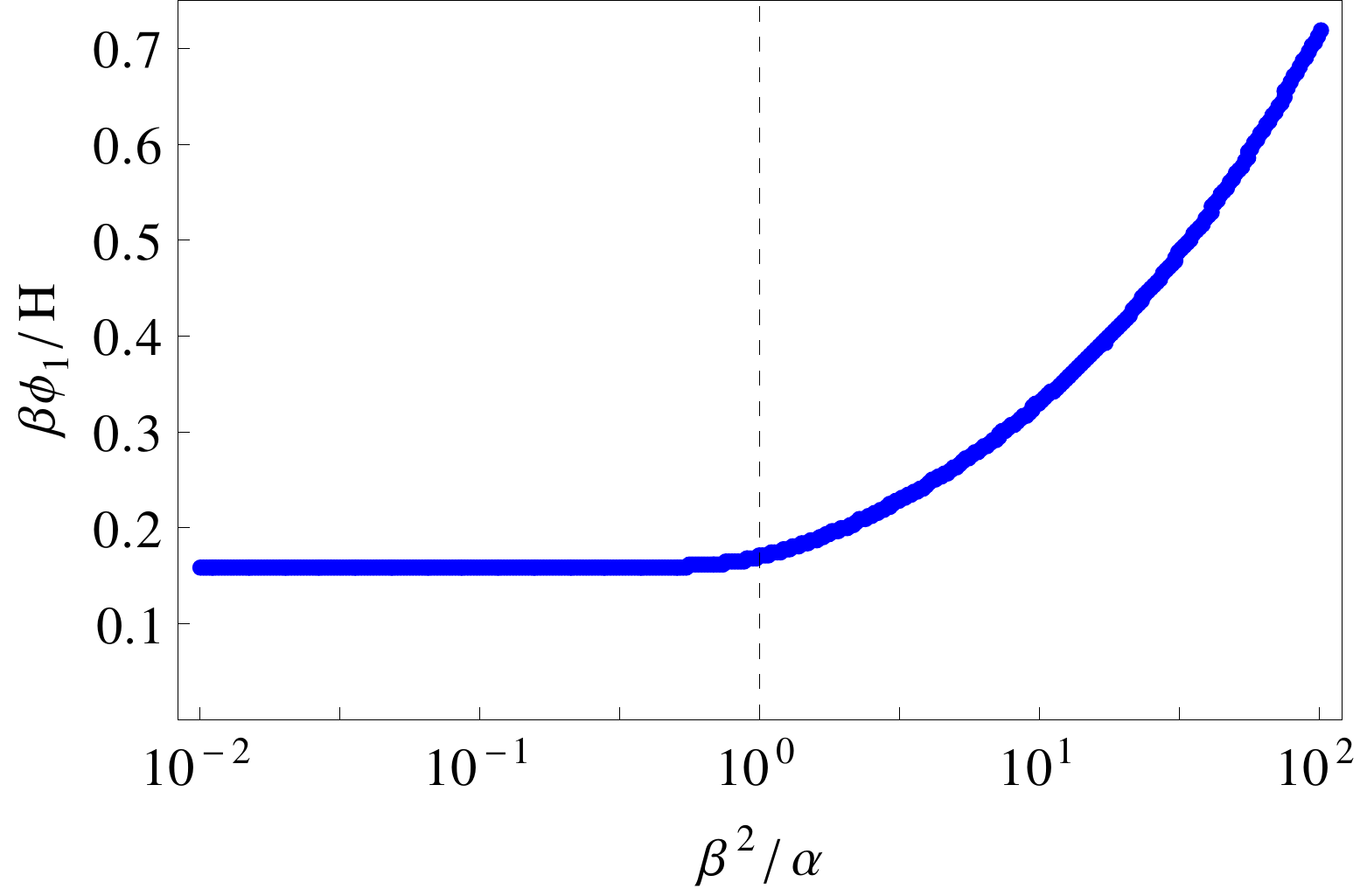} &
\includegraphics[width=0.48\textwidth]{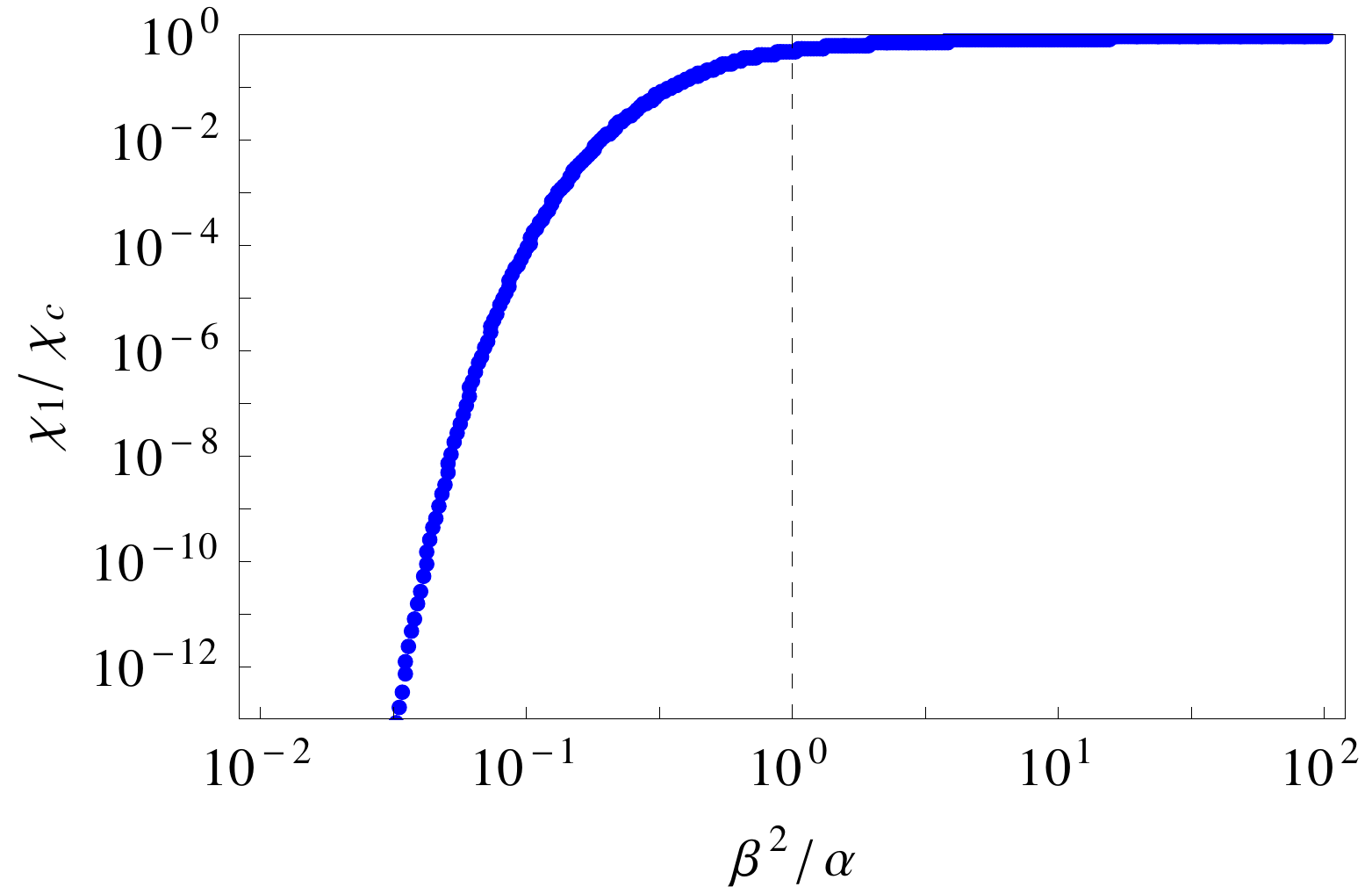}
\end{array}$
  \caption{Values for $\phi_1$ and $\chi_1$ as functions of $\beta^2/\alpha$. We see that for $\beta^2 \ll \alpha$, the value of the preinflaton $\chi_1$ after preinflation is zero up to quantum fluctuations. This implies that for $\beta^2 \ll \alpha$, preinflation generally leads to single-field hilltop inflation with $\phi$ as the only inflaton. Only for $\beta^2 \gtrsim \alpha$, the preinflaton can have an influence during the observable last 60 e-folds, in which case inflation should be studied as a two-field model using e.g.\ the $\delta N$ formalism (see section~\ref{sec:numericalResults}).}
  \label{fig:phi1chi1}
\end{figure}

Eq.~\eqref{eq:phiDiff2} still depends on the point $\Del_0$ where the diffusion region is entered. This is determined from the condition that the diffusion region should not be exited immediately (otherwise, the field will travel at the boundary between the two regions):
\begin{align}
 \left| \frac{ \partial \phi_\text{b}^2 }{ \partial \Del } \right|_{\Del_0} \, = \,  \left( \frac{\Hubble}{2\pi} \right)^2 \frac{2}{\beta^2 \Del_0^3} \, \stackrel{!}{\geq} \, \left| \frac{ \partial \phi_{\text{diff}}^2 }{ \partial \Del } \right|_{\Del_0} \, = \, \left( \frac{\Hubble}{2\pi} \right)^2 \frac{1}{4\alpha ( \Del_0 + 1 )}.
\end{align}
The field stops travelling near the boundary when the inequality starts to be satisfied:\footnote{For arbitrary initial conditions, the fields could enter the diffusion region for any value $\Del \leq \Del_0$. However, when we assume that the fields approach along a trajectory where $\chi \gg \chi_c$ and $\phi \simeq 0$, we know it will move along the boundary until it enters the region at the largest possible value $\Del = \Del_0$.}
\begin{align}
 \frac{ \Del_0^3 }{ 1 + \Del_0 } \, = \, \frac{ 8\alpha }{ \beta^2 }. \label{eq:Del0}
\end{align}
The fields then travel along the trajectory $\phi_\text{diff}(\Del)$ until they exit the diffusion region at $\Del_1$ where $\phi_\text{diff}(\Del)$ and the diffusion boundary $\phi_\text{b}(\Del)$ intersect again:
\begin{align}
 \phi_\text{b}^2(\Del_1) \, = \, \left( \frac{\Hubble}{2\pi} \right)^2 \frac{1}{\beta^2 \Del_1^2} \, \stackrel{!}{=} \, \phi_{\text{diff}}^2(\Del_1) 
 \, = \, \left( \frac{\Hubble}{2\pi} \right)^2 \left[  \frac{1}{\beta^2 \Del_0^2}  +  \frac{1}{4\alpha} \ln \left( \frac{ \Del_0 + 1 }{ \Del_1 + 1 } \right)  \right].
\end{align}
We find an equation to determine $\Del_1$ from $\Del_0$:
\begin{align}
 \frac{1}{\Del_1^2} - \frac{1}{\Del_0^2} + \frac{ \beta^2 }{ 4\alpha } \ln \left( \frac{ \Del_1 + 1 }{ \Del_0 + 1 } \right) \, = \, 0. \label{eq:Del1}
\end{align}
Eqs.~\eqref{eq:Del0} and \eqref{eq:Del1} can be numerically solved for $\Delta_0$ and $\Delta_1$ for any given value of $\frac{\beta^2}{\alpha}$. The result for $\Delta_1$ can then be translated to $\chi_1$ and $\phi_1$ using eqs.~\eqref{eq:defDel} and \eqref{eq:phiB}.\footnote{Note that $\chi_1$ also has a lower bound due to quantum fluctuations. The diffusion region for $\chi$ is given by $\chi < \chi_{\text{b}} = \frac{\Hubble}{2\pi\alpha}$. Below this value, $\chi$ is no longer decaying exponentially but instead behaves as a random variable of order $\chi_\text{b}$. However, $\chi_\text{b}$ is so small that it is negligible for the dynamics during inflation, so this lower bound has no practical relevance in our model.} The results for $\phi_1$ and $\chi_1$ are shown in fig.~\ref{fig:phi1chi1}.

We find that for $\frac{\beta^2}{\alpha} \ll 1$, we start with $\chi_1 \sim \frac{\Hubble}{2\pi \alpha}$ and $\phi_1 \sim \frac{\Hubble}{2\pi \beta}$. The classical equations of motion only drive $\chi$ to smaller values, so it remains small, while $\phi$ is growing over time. As a consequence, $\phi$ is the only relevant field at horizon crossing, and we recover the single-field hilltop inflation limit with $\phi$ as the inflaton.

For larger values of $\frac{\beta^2}{\alpha}$, we can initially have $\chi_1 \sim \chi_\text{crit} \gg \phi_1$. In this case the matter field $\chi$ can have an effect on the spectrum of primordial perturbations. We study this effect numerically in the next sections using the $\delta N$ formalism.

\section{Observables and the $\delta N$ formalism}
\label{sec:deltaN}

To extract predictions for the spectral index $\ns$, the non-Gaussianity parameter $\fnl$ and the amplitude of primordial curvature perturbations $\As$, we employ the $\delta N$ formalism~\cite{Starobinsky:1986fxa, Sasaki:1995aw, Lyth:2005fi}.

The $\delta N$ formalism uses the separate universe approximation, which treats points in the universe that are causally disconnected as being part of independent Friedmann-Robertson-Walker (FRW) spacetimes, with dynamics determined by the local energy density and pressure (see for example~\cite{Lyth, Wands:2000dp}). 
The curvature perturbation evaluated on hypersurfaces of uniform density between two such points, $\zeta$, can then be expressed in terms of the difference of expansion between them, starting from a common initial flat hypersurface labelled $*$, to a final hypersurface of uniform density labelled $f$,
\be
\zeta = \delta N^\text{f}_*.
\en 
In order to extract predictions for the primordial curvature perturbation, the flat hypersurface $*$ must be defined at the time when observable scales exited the cosmological horizon. Moreover, the final hypersurface $f$ must be taken at some much later time when the dynamics are adiabatic and $\zeta$ is conserved. This generally happens after a reheating phase during which the energy is transferred from the inflationary degrees of freedom to a single fluid.

If slow-roll is a good approximation at horizon exit, $\zeta$ 
is completely determined by
the field perturbations $\delta\varphi^i_*$ at that time,
\be
\zeta=\delta N_*^\text{f}(\delta\varphi^i_*).
\en
The field perturbations $\delta\varphi^i_*$ are extremely close to 
Gaussian~\cite{Gaussian}, and if their amplitude 
is sufficiently small, the statistics of 
the curvature perturbation can be determined in a 
simple manner by Taylor expanding
\be
\delta N^\text{f}_* \approx \frac{\partial N^\text{f}_*}{\partial \varphi^{i}_*} \delta\varphi^{i}_* + \frac{1}{2}\frac{\partial^2 N^\text{f}_*}{\partial \varphi^{i}_*\varphi^{j}_*} \delta\varphi^{i}_*\delta\varphi^{j}_*\,,
\label{eq2}
\en
where here and from here on we employ the summation convention, and subsequently we 
will employ the notation in common use, $N_{i}={\partial N^\text{f}_*}/{\partial \varphi^{i}_*}$.

One then finds that the amplitude of the power spectrum
is given by 
\be
\As = N_{i}N_{i}\frac{H_*^2}{4\pi^2}\,,
\label{eq:ampl}
\en 
and the spectral index $\ns$  
by~\cite{Sasaki:1995aw,largeNG1b} 
\be
\ns =1   - 2\epsilon_* + \frac{2}{H_*}\frac{\dot{\varphi}^{i}_*N_{j}N_{ij}}{N_{i}N_{i}} \,.
\label{eq:spectrum}
\en
The non-Gaussianity of the perturbation is characterised by the reduced bispectrum $\fnl$, which is given by~\cite{Lyth:2005fi}
\be
 \fnl = \frac{5}{6}\frac{N_{i}N_{j}N_{ij}}{(N_{i}N_{i})^2}\,.
 \label{eq:fnl}
\en

The current observational bounds on the observables at $68\%$ confidence level are~\cite{Ade:2013zuv}
\be 
 10^9\times \As = 2.1886^{+0.0532}_{-0.0583}\,,
 \label{obsAs}
\en
for the amplitude of the perturbations,
\be
 \ns &=& 0.9603\pm0.0073\,,
 \label{obsns}
 \en
for the spectral index and~\cite{Ade:2013ydc}
\be
 \fnl &=& 2.7\pm5.8\,,
 \label{obsfnl}
 \en
 for the reduced bispectrum.

\section{Numerical results and discussion}
\label{sec:numericalResults}

\subsection{Numerical method}
We studied the inflationary phase using a numerical implementation of the $\delta N$ formalism. To this end, we evolved the homogeneous FRW equations of motion
\begin{eqnarray}
\ddot{\phi} + 3H\dot{\phi} + \frac{\partial V}{\partial \phi} &=& 0\,, \notag\\
\ddot{\chi} + 3H\dot{\chi} + \frac{\partial V}{\partial \chi} &=& 0\,,
\label{homeom}
\end{eqnarray}
where $H$ is the Hubble rate
\begin{equation}
\mpl^2 H^2 = \frac{\rho}{3} = \frac{V}{3}+\frac{\dot{\phi}^2}{6} + \frac{\dot{\chi}^2}{6}\,.
\label{hubble}
\end{equation}
To compute the observables~\eqref{eq:ampl}-\eqref{eq:fnl}, we need to estimate the derivatives $N_{i}$ and $N_{ij}$. This is done by numerically integrating the system~\eqref{homeom} from the field values $\phi_*$ and $\chi_*$ at horizon crossing ($60$ e-folds before the end of inflation), with the derivatives fixed so as to satisfy the slow-roll conditions. The system is integrated until the end of inflation when $\epsilon=1$ and the final Hubble rate $H_\text{f}$ is recorded. After varying the initial conditions at horizon crossing, we restart the integration and end it when $H=H_\text{f}$. We repeat the process until we have enough points to build up a finite difference estimation of the derivatives of $N(\phi_*,\chi_*)$.

Ideally, the final hypersurface of constant energy density $H_\text{f}$ should be chosen when the evolution of the universe is adiabatic. In our scenario, at the end of inflation all the energy density is stored in the fields $\phi$ and $\chi$ and is subsequently transferred to non-inflationary degrees of freedom, possibly generating a further evolution of the curvature perturbation $\zeta$. We will see that in some regions of parameter space, the field trajectory reaches the adiabatic limit before the end of inflation, implying that the super-horizon observables evaluated at that time are conserved. For other parameters, however, this is not the case. In that case, one needs to study the evolution of the curvature perturbation during the following reheating phase (up to the point where the perturbations become adiabatic), which is beyond the scope of this paper.

The parameter space of the model~\eqref{vinf}, including the initial field and derivative values, is $9$-dimensional. The initial field values $\phi_{\rm i}$ and $\chi_{\rm i}$ are set at the exit of the diffusion region, given by eqs.~\eqref{eq:Del0} and~\eqref{eq:Del1} as explained in section~\ref{sec:ics} and their derivatives are fixed using the slow-roll approximation. The parameter $\Lambda$ is rescaled to match the amplitude of perturbations $\As$. Such rescaling affects the calculation of the initial field values and is therefore calculated iteratively. We spanned the following region of parameter space:
\begin{align}
\lambda\mpl \in \{10^{-3},  1\} \,,\quad~~
10^{-5}\leq \, \alpha \, \leq 10^{-2} \, , \quad~~
10^{-3}\leq \, \beta \, \leq 4\times10^{-2}\,.
\label{params}
\end{align}
All plots are shown for $\mu = 10^{-5/2}\mpl$. The effects of different $\mu$ will be discussed below. The non-Gaussianity parameter $\fnl$ turns out to be extremely small, with $\lvert \fnl \rvert < 0.1 $, so it is not shown.

\begin{figure}[tbph]
\centering
$\begin{array}{c}
\includegraphics[width=0.8\textwidth]{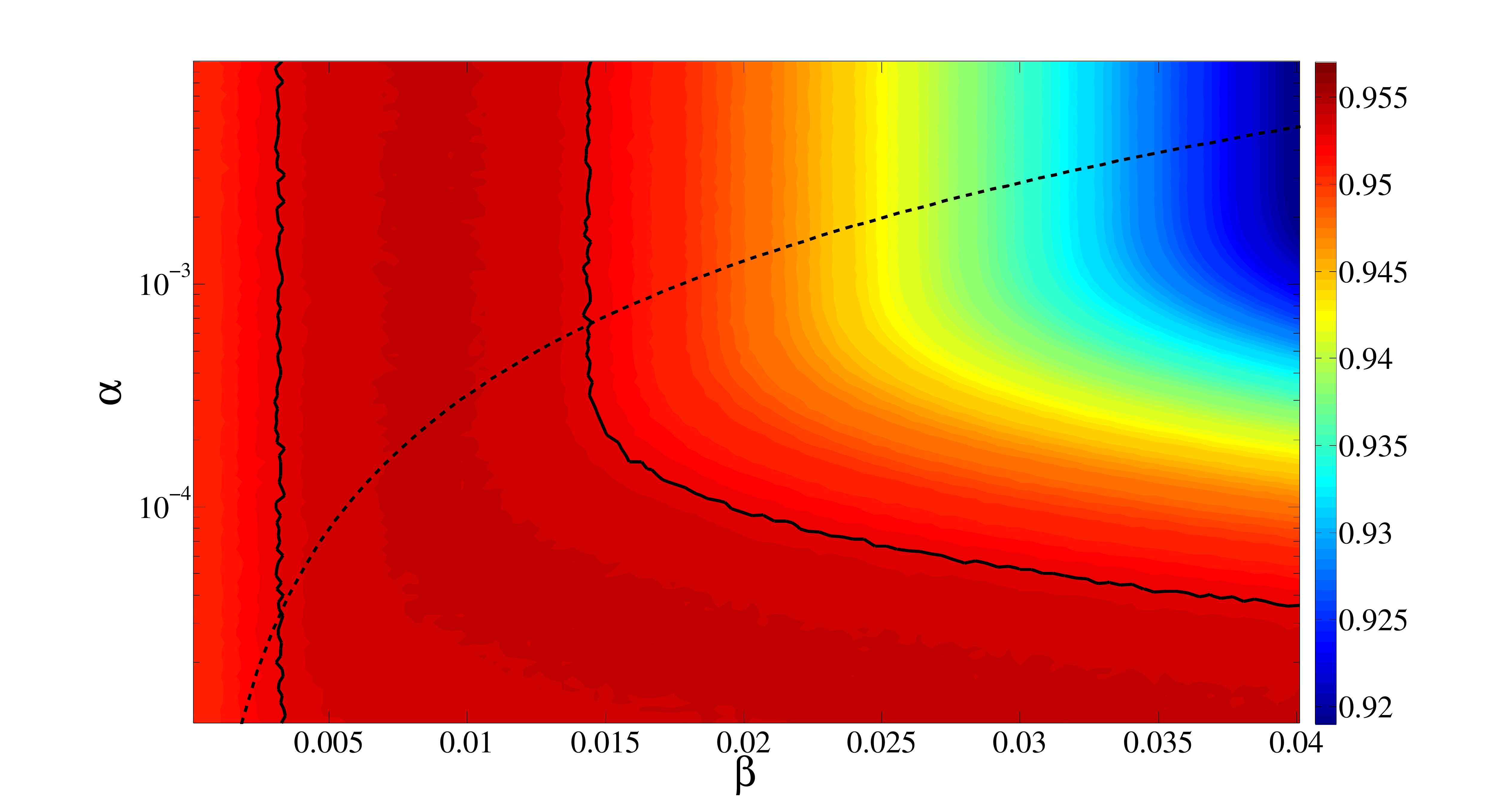} \\
\includegraphics[width=0.8\textwidth]{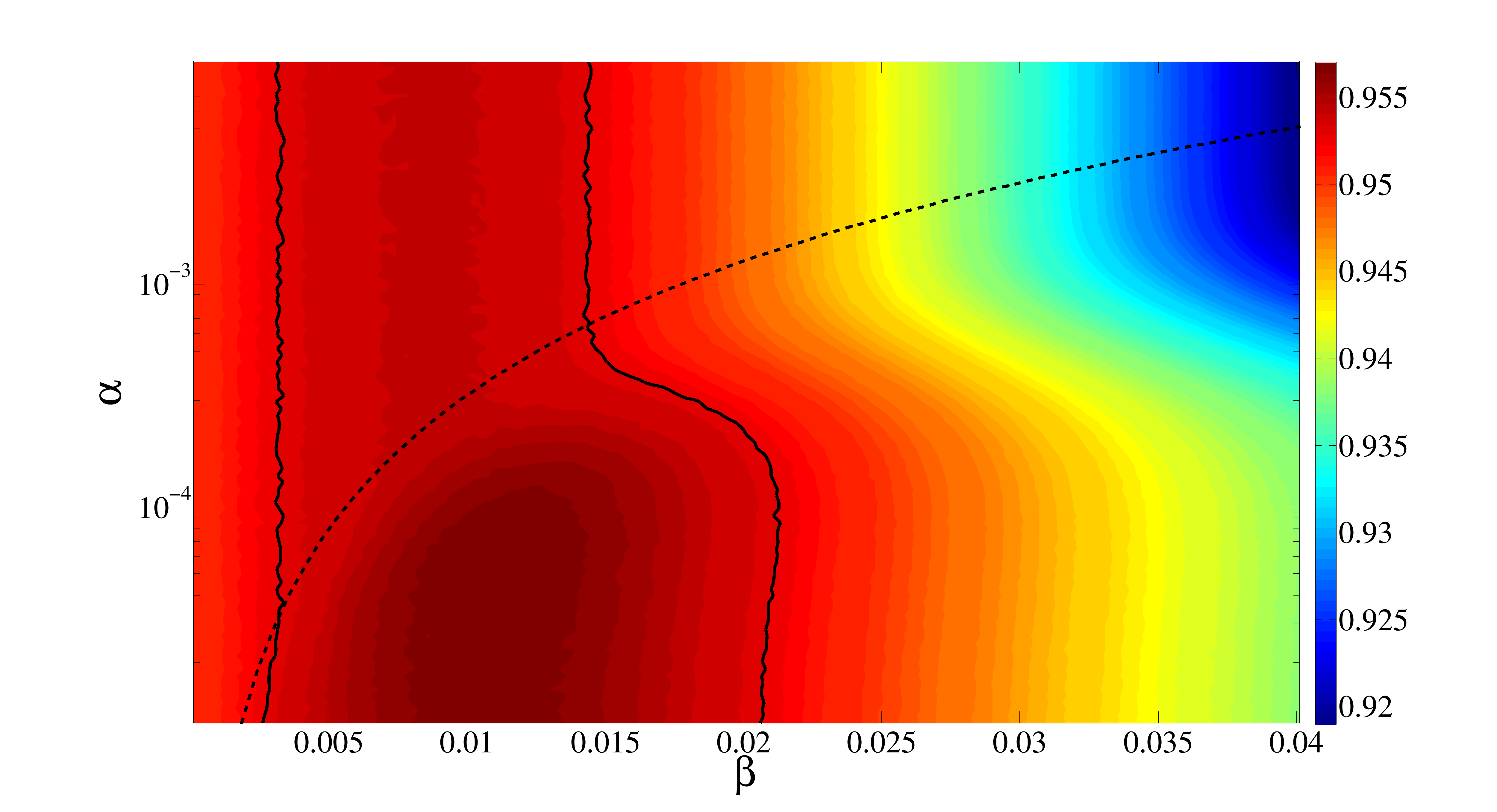}
\end{array}$ 
  \caption{Contour plot of the spectral index $\ns$ as a function of $\alpha$ and $\beta$, with $\lambda\mpl=10^{-3}$ (top) and $\lambda\mpl=1$ (bottom). $\mu=\sqrt{10^{-5}}\mpl$ and the initial field values are given by eqs.~\eqref{eq:Del0} and~\eqref{eq:Del1}. The black contour lines correspond to $\ns=0.953$ and values contained within the lines are in agreement with the $68\%$ C.L. bounds of Planck. The black dashed line corresponds to $\alpha/\beta^2=1.4$.}
  \label{fig:ns}
\end{figure}

\begin{figure}[tbph]
\centering
$\begin{array}{c}
\includegraphics[width=0.8\textwidth]{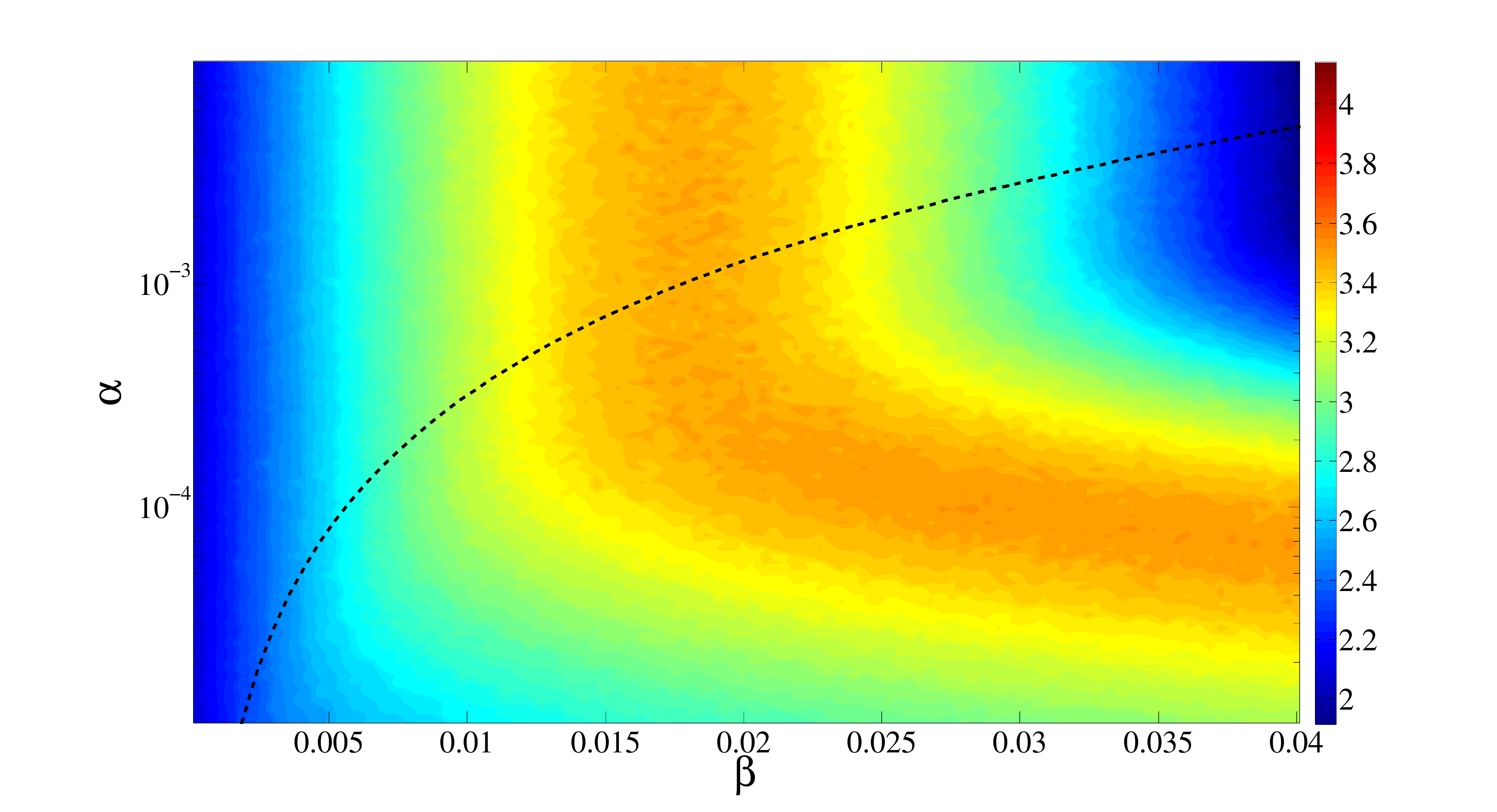} \\
\includegraphics[width=0.8\textwidth]{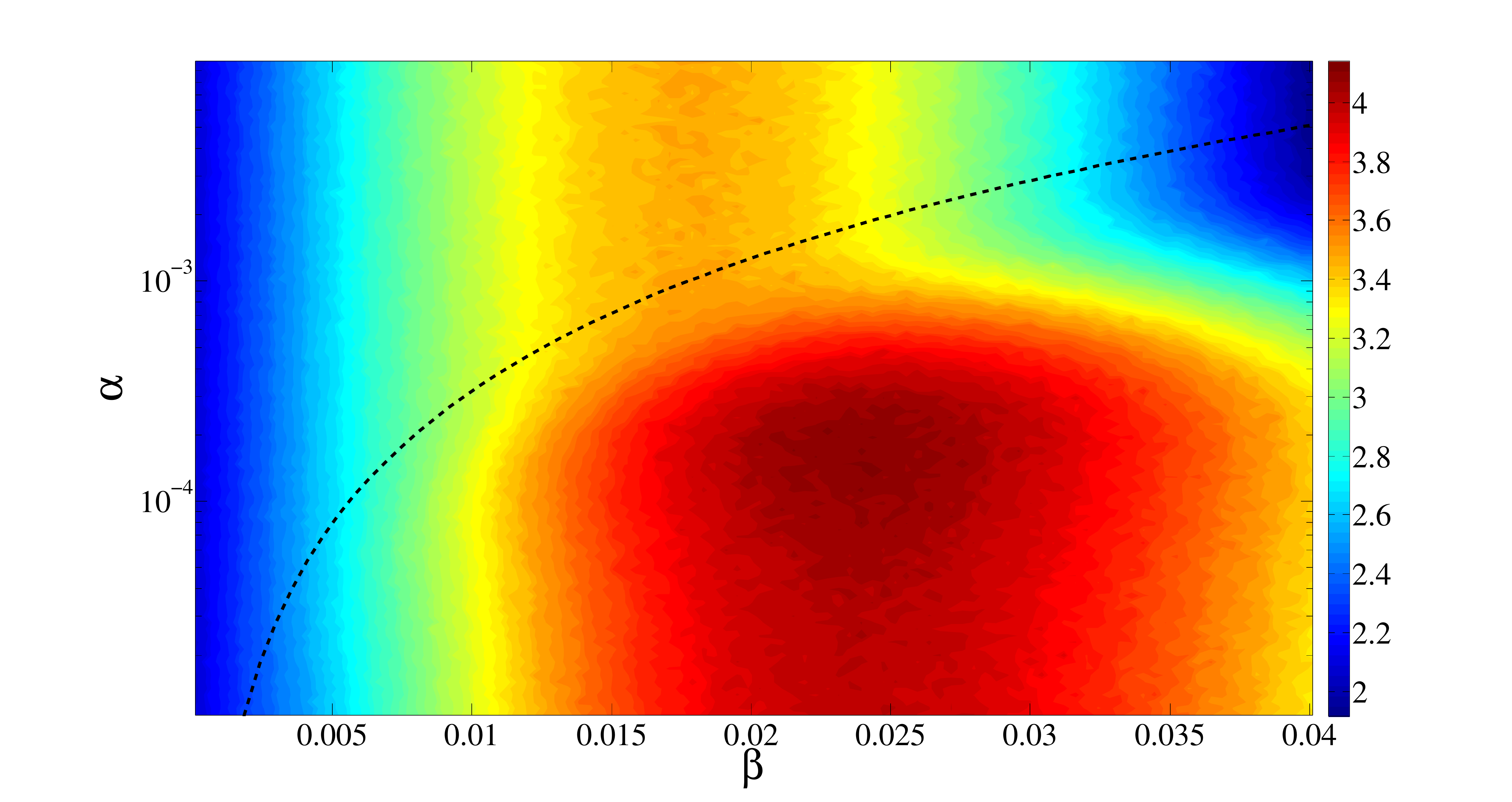}
\end{array}$ 
  \caption{Contour plot of $10^{24}\times\Lambda^4/\mpl^4$ as a function of $\alpha$ and $\beta$, with $\lambda\mpl=10^{-3}$ (top) and $\lambda\mpl=1$ (bottom). $\mu=\sqrt{10^{-5}}\mpl$ and the initial field values are given by eqs.~\eqref{eq:Del0} and~\eqref{eq:Del1}, after the rescaling to match the amplitude of primordial perturbations. The black dashed line corresponds to $\alpha/\beta^2=1.4$.}
  \label{fig:vnot}
\end{figure}

\begin{figure}[tbph]
  \centering
$\begin{array}{cc}
\includegraphics[width=0.48\textwidth]{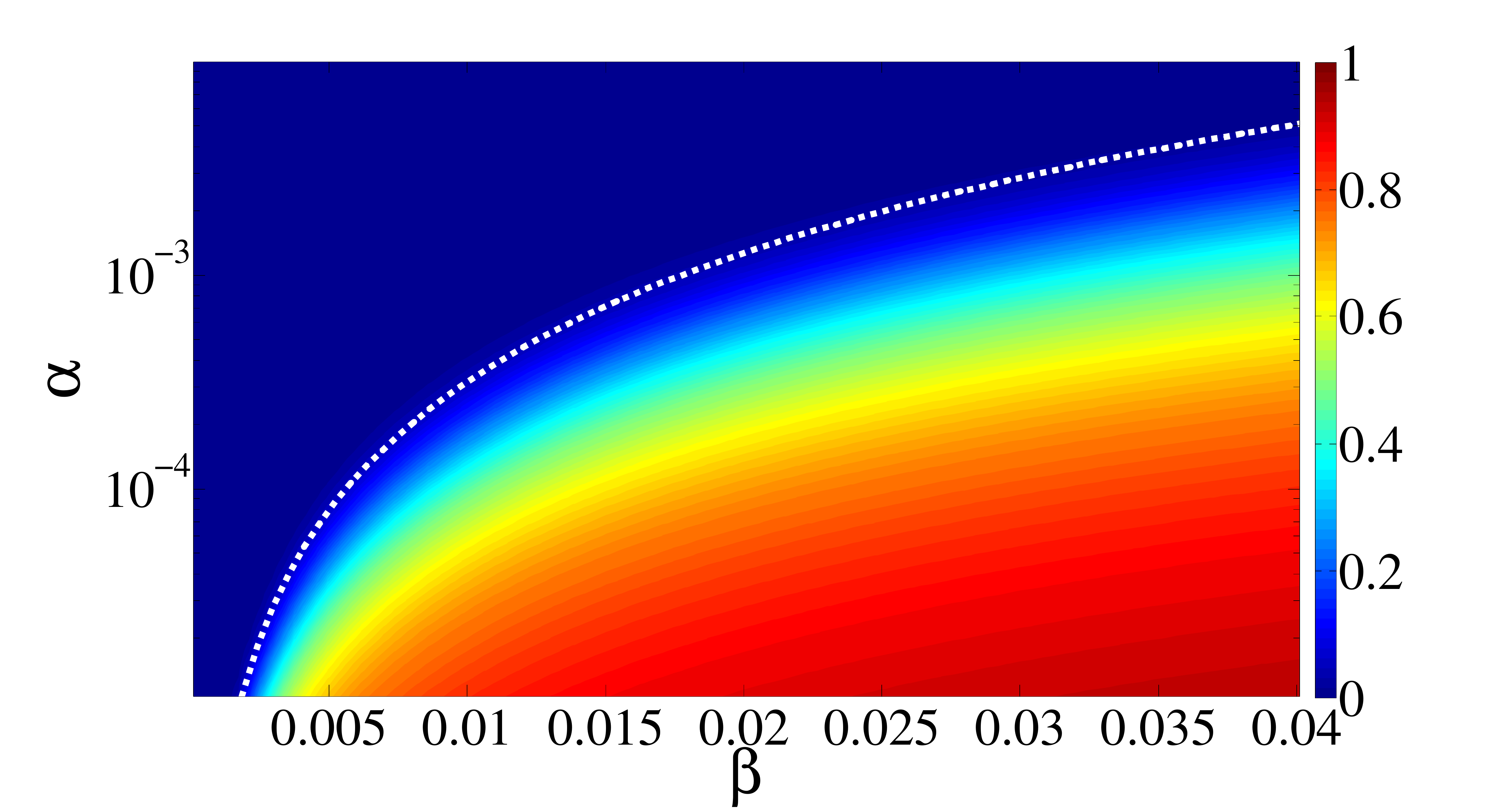} &
\includegraphics[width=0.48\textwidth]{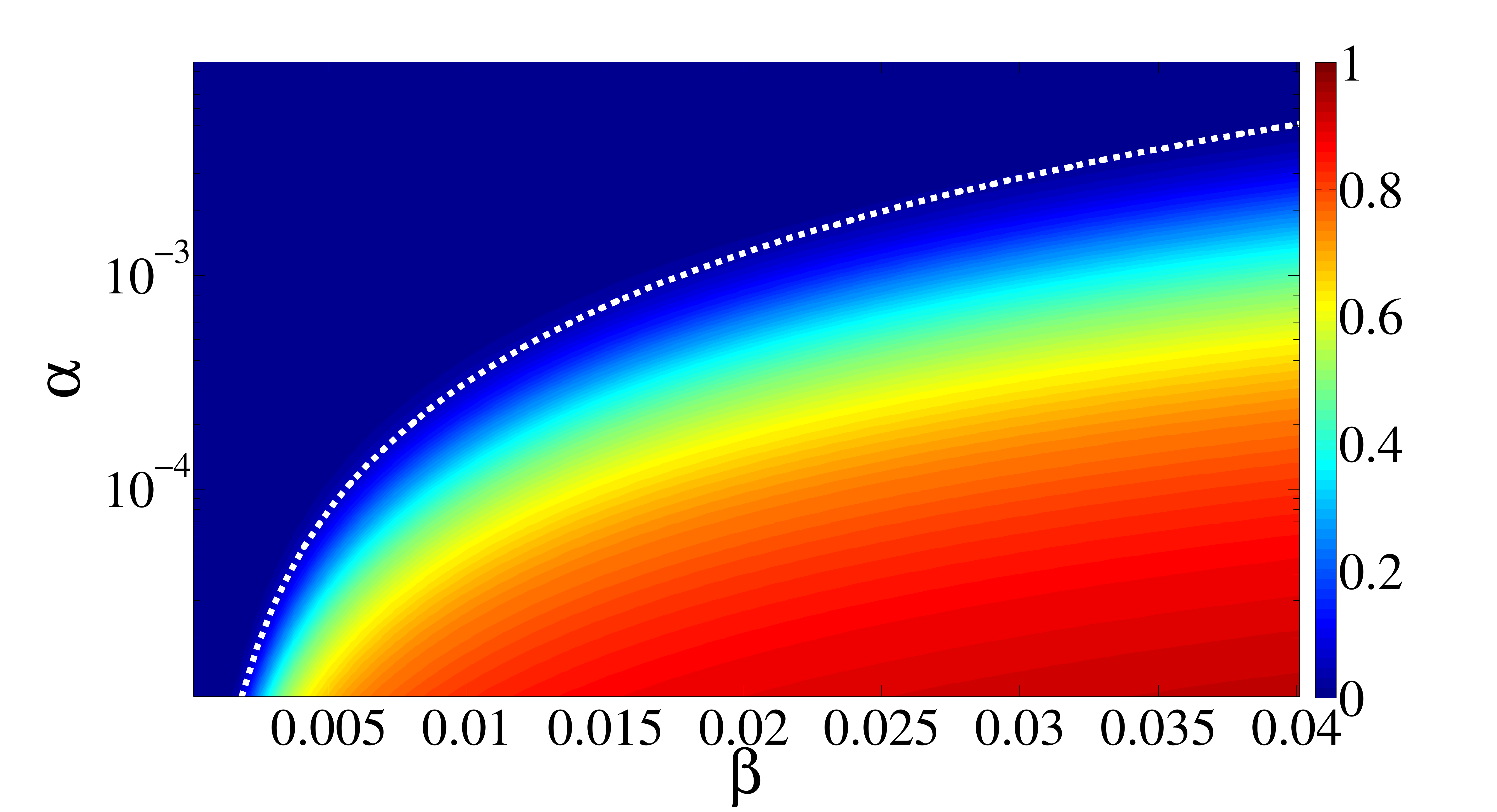} 
\end{array}$
  \caption{Contour plot of $\chi/\chi_{\rm crit}$ at horizon crossing (left) and at the end of inflation (right) as a function of $\alpha$ and $\beta$, with $\lambda\mpl=10^{-3}$ and initial field values given by eqs.~\eqref{eq:Del0} and~\eqref{eq:Del1}. The white dashed line corresponds to $\alpha/\beta^2=1.4$. Above this line, $\chi_*/\chi_{\rm crit}$ is negligible and the model effectively reduces to single field hilltop inflation. Below the dashed line, $\chi_*/\chi_{\rm crit} \sim 1$ and the preinflaton field has to be taken into account. As can be seen by comparing the two plots, $\chi$ is nearly constant during inflation, thus its only effect during inflation is to shift $\beta \rightarrow \beta_{\rm eff}$ and $\mu\rightarrow\mu_{\rm eff}$ as defined in eq.~\eqref{eq:lambdaMuEff}. However, at the end of inflation the dynamics are not adiabatic yet and reheating may affect the primordial observables.}
  \label{fig:chihe}
\end{figure}

\begin{figure}[tbph]
  \centering
$\begin{array}{cc}
\includegraphics[width=0.48\textwidth]{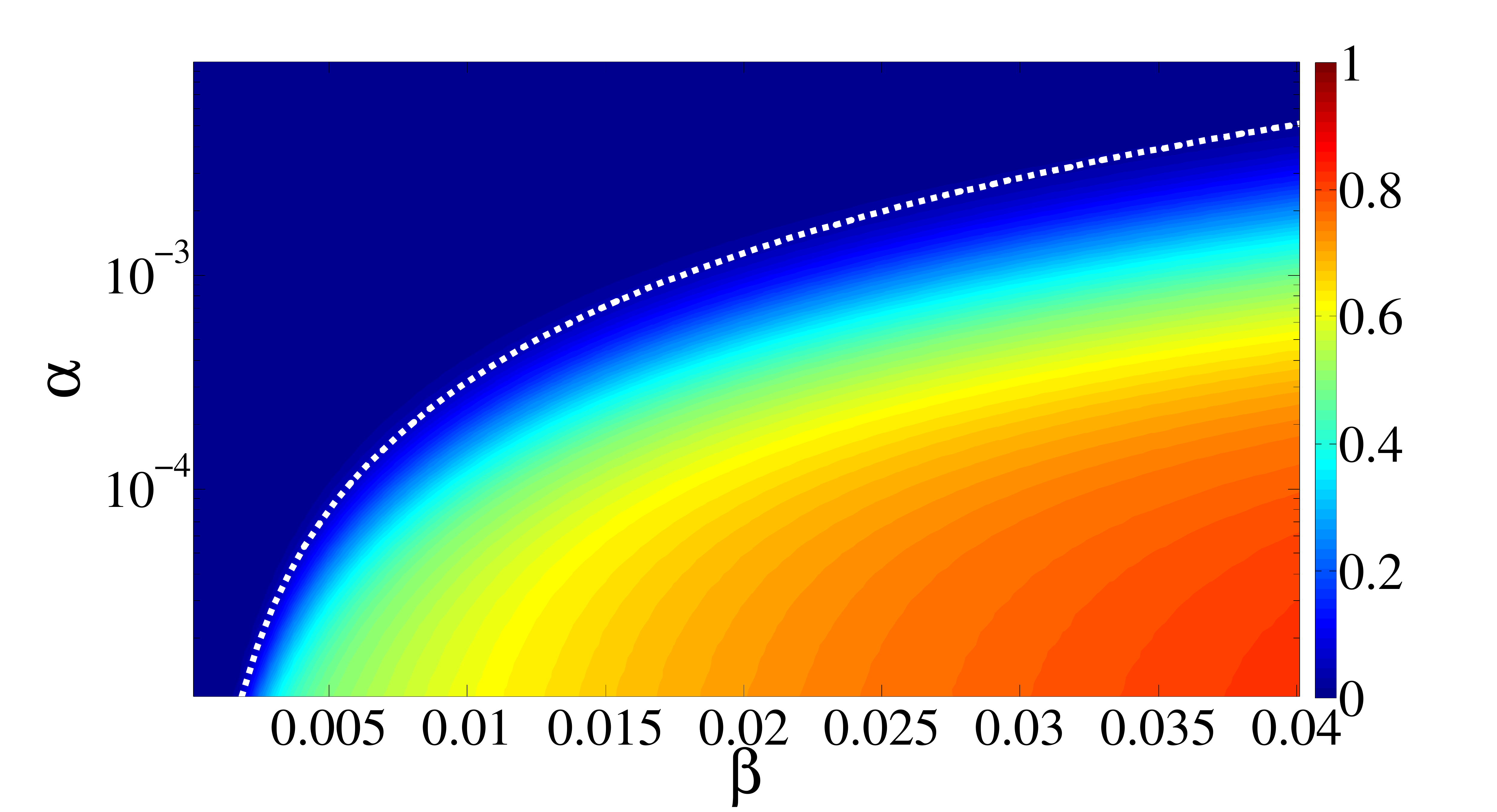} &
\includegraphics[width=0.48\textwidth]{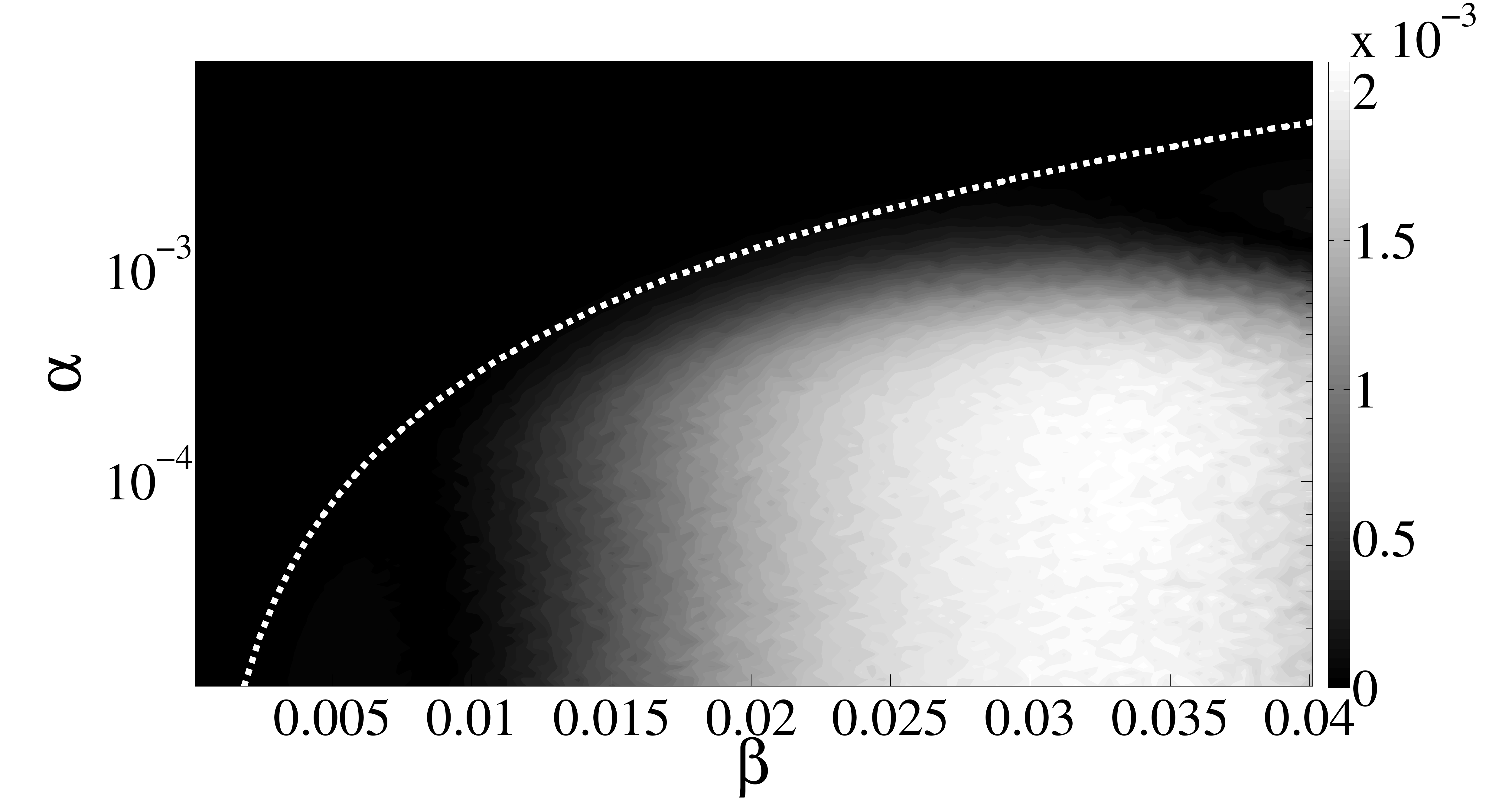} 
\end{array}$
  \caption{Contour plot of $\chi/\chi_{\rm crit}$ at horizon crossing (left) and at the end of inflation (right) as a function of $\alpha$ and $\beta$, with $\lambda\mpl=1$ and initial field values given by eqs.~\eqref{eq:Del0} and~\eqref{eq:Del1}. The white dashed line corresponds to $\alpha/\beta^2=1.4$. Above this line, $\chi_*/\chi_{\rm crit}$ is negligible and the model effectively reduces to single field hilltop inflation. Below the dashed line, $\chi_*/\chi_{\rm crit} \sim 1$ and the preinflaton field has to be taken into account. As can be seen in the right plot, at the end of inflation $\chi$ is oscillating about the minimum with maximal amplitude $\chi\simeq10^{-3}\chi_{\rm crit}$, suggesting that the universe has reached the adiabatic limit.}
  \label{fig:chihe1}
\end{figure}

The predictions for the spectral index $\ns$ and the vacuum energy $\Lambda^4$ are shown in figs.~\ref{fig:ns} and \ref{fig:vnot}, both for $\lambda \mpl = 10^{-3}$ and for $\lambda \mpl = 1$. We find that for $\alpha/\beta^2 > \alphaBetaThr$ (above the dashed line), the results are identical to single-field hilltop inflation. In particular, they are independent of the preinflaton couplings $\alpha$ and $\lambda$.

Below the dashed line, where $\alpha/\beta^2 < \alphaBetaThr$, we find that the predictions for $\ns$ and $\Lambda$ are changed, and that they depend on the coupling $\lambda$. There are two qualitatively different regimes, depending on whether $\lambda$ is small or large. For small $\lambda$, $\chi$ is effectively a constant background field during slow-roll inflation in $\phi$, whereas for large $\lambda$, inflation proceeds along a non-trivial multi-field trajectory. Exactly what constitutes a large or small coupling $\lambda$ depends on the value of $\mu$. For $\mu = 10^{-5/2} \mpl$, which is the value chosen for our plots, $\lambda \mpl = 10^{-3}$ is in the small coupling regime, whereas $\lambda \mpl = 1$ is in the large coupling regime.

We thus find three qualitatively different regimes:
\begin{enumerate}
 \item The limit of \textbf{single-field hilltop inflation} ($\alpha/\beta^2 > \alphaBetaThr$), where $\chi \sim 0$ and inflation happens as single-field inflation in $\phi$.
 \item The limit of \textbf{quasi-single-field hilltop inflation} ($\alpha/\beta^2 < \alphaBetaThr$, small $\lambda$), where $\chi \sim \text{constant}$ and slow-roll inflation happens as single-field inflation in $\phi$, with the inflaton potential $V(\phi)$ modified by the nearly constant background field $\chi$.
 \item Non-trivial \textbf{multi-field inflation} ($\alpha/\beta^2 < \alphaBetaThr$, large $\lambda$), where both $\chi$ and $\phi$ are dynamic fields during slow-roll inflation.
\end{enumerate}

\subsection{Discussion of the different regimes}

\subsubsection*{Single-field limit for $\alpha/\beta^2 > \alphaBetaThr$}

As we discussed in section~\ref{sec:ics}, the initial conditions for $\alpha/\beta^2 \gg 1$ converge towards $\phi_\text{i} \rightarrow 0$, $\chi_\text{i} \rightarrow 0$. This means that inflation happens along the single-field hilltop inflation trajectory with $\chi \simeq 0$. This can also be seen in figs.~\ref{fig:chihe} and~\ref{fig:chihe1}, where we plot the numerical results for the preinflaton field value $\chi_*$ at horizon crossing and $\chi_\text{e}$ at the end of slow-roll inflation. For $\alpha/\beta^2 > \alphaBetaThr$, the preinflaton is negligible already at horizon crossing. For this reason, the predictions in this limit do not depend on the preinflaton couplings $\alpha$ and $\lambda$.

As in single-field hilltop inflation, $\ns$ does not depend on $\mu$ in this region, and $\Lambda^4 \propto \mu^4$.

\subsubsection*{Small $\lambda$ regime for $\alpha/\beta^2 < \alphaBetaThr$}

When $\alpha/\beta^2$ is small, the initial preinflaton field value $\chi_\text{i}$ is large enough for $\chi$ to have an impact during inflation. Therefore, the dynamics during inflation depend on the preinflaton coupling $\lambda$.

For small coupling $\lambda$, the preinflaton potential is extremely flat (note that $\alpha < \alphaBetaThr \,\beta^2$ implies that the preinflaton mass term is small, so a sizeable potential for $\chi$ can only be generated from the coupling $\lambda$). We therefore expect that for sufficiently small $\lambda$, $\chi$ is nearly constant throughout slow-roll inflation: $\chi(t) \simeq \chi_* \simeq \chi_\text{e}$. This is well confirmed by our numerical results for $\chi_*$ and $\chi_\text{e}$ which are plotted in fig.~\ref{fig:chihe} for $\lambda \mpl = 10^{-3}$.

In this limit for small $\lambda$, the slow-roll dynamics are then given by single-field slow-roll inflation in $\phi$ in the presence of a constant background field $\chi_*$, which leads to an inflaton potential
\begin{align}
 V(\phi) \, &\simeq \, \Lambda^4\left( 1 - 2\frac{\phi^4}{\mu^4} - \frac{\beta}{2\mpl^2} \phi^2 \right) + \frac{1}{2} \lambda^2 \chi_*^2 \phi^4 + \frac{1}{2} \lambda^2 \chi_*^4 \phi^2 \notag\\
 &= \, \Lambda^4 \left( 1 - 2\frac{\phi^4}{\mu_\text{eff}^4} - \frac{\beta_\text{eff}}{2\mpl^2} \phi^2 \right),
\end{align}
with
\begin{align}
 \beta_\text{eff} \, = \, \beta\left( 1 - \frac{\chi_*^4}{\chi_\text{crit}^4} \right), \quad \mu_\text{eff} \, = \, \mu \left( 1 - \frac{\lambda^2 \chi_*^2 \mu^4}{4\Lambda^4} \right)^{-1/4}. \label{eq:lambdaMuEff}
\end{align}
We therefore recover the single-field hilltop inflation limit with redefined parameters $\beta \rightarrow \beta_\text{eff}$ and $\mu \rightarrow \mu_\text{eff}$. This is confirmed by our numeric results for $\ns$ and $\Lambda^4$ in figs.~\ref{fig:ns} and \ref{fig:vnot}, where for $\lambda \mpl = 10^{-3}$ the region below the dashed line reproduces the single-field predictions, just stretched along the $\beta$ direction with a stretching factor $\beta \rightarrow \beta(1 - \chi_*^4/\chi_\text{crit}^4)$.

This regime once again effectively reduces to single-field hilltop inflation (albeit with modified parameters $\beta_\text{eff}$ and $\mu_\text{eff}$), and we thus know that $\ns$ is independent of $\mu$, while $\Lambda^4 \propto \mu_\text{eff}^4$, where $\mu_\text{eff}$ as a function of $\mu$ is given in eq.~\eqref{eq:lambdaMuEff}.

Note, however, that we have calculated the spectrum of perturbations at the end of slow-roll inflation. As the preinflaton field $\chi$ has not yet reached the minimum at that time, the universe has not reached the adiabatic limit yet, and the spectrum of primordial curvature perturbations could still be changed during the reheating period after inflation (see e.g.\ \cite{spectrumFromReheating} for perturbative reheating and ~\cite{preheating} for non-equilibrium effects). As the reheating process is model dependent, these effects cannot be addressed in our general framework and must be checked for each specific model.

\subsubsection*{Large $\lambda$ regime for $\alpha/\beta^2 < \alphaBetaThr$}

When $\lambda$ is large, the preinflaton $\chi$ gets mass and self-coupling terms $\frac{1}{2} \lambda^2 \phi^2 \chi^2 (\phi^2 + \chi^2)$. As they are proportional to $\phi^2$ and $\phi^4$, these couplings start out small, so $\chi$ does not roll towards zero before horizon crossing (if it did, we would end up in the single-field hilltop inflation limit discussed above). However, when $\phi$ grows to larger values during inflation, the potential for $\chi$ becomes increasingly steep, and $\chi \rightarrow 0$. This broad picture is confirmed by our numerical result for $\chi_*$ and $\chi_\text{e}$, which is plotted in fig.~\ref{fig:chihe1} for $\lambda \mpl = 1$: we see that below the dashed line, we have $\chi_* \sim \chi_\text{crit}$, but $\chi_\text{e} \ll \chi_\text{crit}$.

In this regime, slow-roll inflation happens along a non-trivial multi-field trajectory, and the predictions depend non-trivially on the model parameters $\mu$ and $\lambda$. In our example, we see that for $\mu = 10^{-2.5} \mpl$, the spectral index $\ns$ can be closer to the central value measured by Planck (for $\alpha \lesssim 10^{-4}, \beta \sim 0.01$). However, we find numerically that for different $\mu$, the deviations can instead lower the spectral index in this region.

If $\lambda$ is sufficiently large, the preinflaton $\chi$ usually reaches the minimum before the end of inflation, so the evolution at the end of inflation should already be adiabatic. Otherwise, the spectrum of curvature perturbations could be affected by reheating in the same way as discussed for the small $\lambda$ regime above.

\subsection{Effect of $\operatorname{Im}(\Phi)$}
\label{sec:imaginaryInflatonEffects}

So far, we have ignored the imaginary inflaton component $\operatorname{Im}(\Phi)$. It has been shown in \cite{Nolde:2013bha} that the imaginary inflaton component can reduce the spectral index $\ns$ and the vacuum energy $\Lambda^4$ in supersymmetric hilltop inflation, depending on the inflaton mass parameter $\beta$. For small $\beta \ll 10^{-2}/(2p)$, this effect is negligible, whereas for larger $\beta$, the magnitude of the reduction in $\ns$ and $\Lambda^4$ depends on the initial ratio $\operatorname{Im}(\Phi) / \operatorname{Re}(\Phi)$, which is a random variable given by quantum fluctuations.

For the single-field and the quasi-single-field regime, these results are directly applicable, as the last 60 e-folds of inflation are well-described by supersymmetric hilltop inflation with the complex inflaton $\Phi$. Only for the large $\lambda$ multi-field case, the effect of the imaginary inflaton component may be different. In this case, it must generally be included in the numerical calculation, with the initial condition $\operatorname{Im}(\Phi) / \operatorname{Re}(\Phi)$ as an additional free parameter.

\section{Reheating and leptogenesis from coupling to matter field}
\label{sec:reheating}

Using matter fields for preinflation is not only economical, it also provides an interesting decay channel for reheating. The inflaton-matter coupling used for preinflation allows decays of the inflaton particle $\phi$ into the matter field $\chi$. The matter field can then decay further into Standard Model particles to reheat the universe. As the couplings within the matter sector are related to particle physics observables, this leads to calculable predictions for the reheat temperature and related quantities like the baryon asymmetry.

To illustrate the benefits of using matter fields for preinflation, we discuss an explicit example where the right-handed sneutrino takes the role of the preinflaton.\footnote{
For a discussion of using the sneutrino as the inflaton, see e.g.\ \cite{sneutrinoChaotic,sneutrinoChaotic2} for chaotic inflation or \cite{sneutrinoTribrid} for hybrid inflation.
} For this example, we calculate the reheat temperature and the baryon asymmetry that is generated from out-of-equilibrium decays during reheating. Demanding that the generated baryon asymmetry from nonthermal leptogenesis matches the observed value, we can then constrain the masses of both the inflaton and one of the right-handed neutrinos.

\subsection{Example model: sneutrino preinflation}

As an explicit example, we assume that one of the right-handed sneutrinos $X_i$ takes the role of the preinflaton $\chi$. The superpotential is
\begin{align}
 W \, = \, \Lambda^2 \hat{S} \left( \frac{ 4\hat{\Phi}^4 }{ \mu^4 } - 1 \right)  +  \lambda_i \hat{\Phi}^2 \hat{X}_i^2  +  y_{ji} \hat{L}_{j} \hat{H}_u \hat{X}_i, \label{eq:Wexample}
\end{align}
which is a sum of the inflaton superpotential from eq.~\eqref{wgen} and neutrino Yukawa couplings including the left-handed lepton doublets $\hat{L}_j$ and electroweak up-type Higgs doublet $\hat{H}_u$. The right-handed neutrino Majorana mass term is generated from the vacuum expectation value of the inflaton $\Phi$ after inflation. We assume that $L_j = H_u = 0$ during inflation, so the Yukawa couplings have no effect on inflation.\footnote{The left-handed sleptons and the electroweak Higgs can get Hubble-sized mass terms from the \kahler potential which stabilize them at zero.}

With this superpotential, the inflaton can decay into right-handed sneutrinos and right-handed neutrinos which then continue to decay into left-handed (s)leptons and electroweak up-type Higgs(ino) particles.

A tree-level two-particle decay of the inflaton requires that $m_\phi > 2m_{X_i}$ for at least one of the right-handed neutrino superfields $\hat{X}_i$. In this example, we assume for simplicity that $m_\phi > 2m_{X_i}$ is satisfied for exactly one of the right-handed neutrinos. We drop the index $i$ from now on both on this right-handed neutrino superfield $\hat{X}$ and its superpotential coupling $\lambda$. The calculation can easily be generalized to the more general case where the inflaton can decay into more than one of the $X_i$.

\subsection{Decay rates}

To calculate the reheat temperature and baryon asymmetry, we need to calculate the inflaton decay rate $\Gamma_\phi$ into right-handed (s)neutrinos. We will also calculate the decay rates of right-handed (s)neutrinos into (s)leptons and Higgs(ino) particles, which are assumed to thermalize quickly due to their gauge interactions.

Assuming $m_\phi > 2m_X$ so that the inflaton can decay into pairs of right-handed (s)neutrinos, we find the decay rate for the inflaton $\phi$
\begin{align}
 \Gamma_\phi \, = \, \frac{\lambda}{8\pi} m_\phi m_X \left(  1 + 12 \frac{m_X^2}{m_\phi^2}  \right)\left(  1 - 4 \frac{m_X^2}{m_\phi^2}  \right)^{1/2}
\end{align}
and the decay rates for the right-handed neutrino $\psi_X$ and the two sneutrino components $\chi$, $\chiI$
\begin{align}
 \Gamma_\chi \, = \, \Gamma_{\chiI} \, = \, \Gamma_{\psi_X} \, = \, \frac{ \sum_j \lvert y_{ji} \rvert^2 }{4\pi} m_X.
\end{align}
The calculation of these decay rates is discussed in appendix \ref{appendix:decayRates}.

As reheating happens through the decay chains $\phi \rightarrow \chi + \chi \rightarrow \text{MSSM}$, $\phi \rightarrow \chiI + \chiI \rightarrow \text{MSSM}$ and $\phi \rightarrow \psi_X + \psi_X \rightarrow \text{MSSM}$, the relevant decay rate for the entire decay chain will be the minimum of $\Gamma_\phi$ and $\Gamma_\chi = \Gamma_{\chiI} = \Gamma_{\psi_X}$. In our model, we find the ratio
\begin{align}
 \frac{\Gamma_\phi}{\Gamma_\chi} \, &= \, \frac{ \lambda m_\phi }{ 2 \sum\limits_j \lvert y_{ji} \rvert^2 }\left(  1 + 12 \frac{m_X^2}{m_\phi^2}  \right)\left(  1 - 4 \frac{m_X^2}{m_\phi^2}  \right)^{1/2} \, \sim \, \frac{ \lambda m_\phi }{ 2 \sum\limits_j \lvert y_{ji} \rvert^2 }. \label{eq:decayRateRatio}
\end{align}
For the following discussion, we will assume that $\Gamma_\phi < \Gamma_\chi$, which is generally the case if the neutrino Yukawa couplings are sufficiently large. We will later show that our results are consistent with this assumption (see subsection~\ref{sec:neutrinoMassScale}).

If the Yukawa couplings were very small, so that $\Gamma_\phi > \Gamma_\chi$, the calculation would be identical to \cite{valerieLeptogenesis}, where the $\phi$ field decays quickly and reheating can be studied by only considering the sneutrino and neutrino decays via their Yukawa coupling. However, for our model the case with $\Gamma_\phi < \Gamma_\chi$ is more generic because with our superpotential \eqref{eq:Wexample} the mass $m_\phi$ is smaller than in the model studied in \cite{valerieLeptogenesis}. We therefore focus on the case $\Gamma_\phi < \Gamma_\chi$ for the rest of this paper.

In the simplest case, reheating happens via perturbative inflaton decays and the decay products thermalize quickly due to the efficient gauge interactions. In that case, the reheat temperature and the produced lepton asymmetry can be estimated from the inflaton decay rate $\Gamma_\phi$ and the inflaton and sneutrino masses $m_\phi$ and $m_X$ as outlined below.

For a detailed analysis, one should also consider nonperturbative effects from tachyonic preheating and parametric resonance \cite{preheatingNewInflation}. Rapid thermalization of the decay products is also not guaranteed if some D-flat MSSM directions develop large vacuum expectation values during inflation. The calculation in this section assumes that nonperturbative preheating effects are small and that thermalization happens fast compared to the inflaton decay rate.

\subsection{Inflaton and (s)neutrino masses}

The inflationary vacuum energy $\Lambda^4$ is fixed by the amplitude $\As$ of scalar perturbations. As the numerical calculations in section \ref{sec:numericalResults} have shown, $\Lambda^4$ does not depend very much on $\alpha$ and $\beta$, so we estimate it from the single-field hilltop inflation limit with $\beta=0$, which is
\begin{align}
 \Lambda^4 \, \simeq \, \frac{ 3 \pi^2 \As }{ 16 \Ne^3 } \mu^4 \, \simeq \, \left( 2 \times 10^{-14} \right) \mu^4
\end{align}
for $\Ne \simeq 60$ and $\As = 2.2 \times 10^{-9}$.

The inflaton mass $m_\phi$ and the right-handed sneutrino mass $m_X$ at the global minimum can be calculated in terms of the model parameters $\mu$ and $\lambda$:
\begin{subequations}
\begin{align}
 m_\phi \, &= \, 16\frac{\Lambda^2 \braket{\Phi}^3}{\mu^4} \, = \, 2^{5/2} \frac{\Lambda^2}{\mu} \, \simeq \, \left( \frac{ 6 \pi^2 \As }{ \Ne^3 } \right)^{\frac12} \mu \, = \, (8\times10^{-7}) \,\mu, \label{eq:mPhi}\\
 m_X \, &= \, 2\lambda \braket{\Phi}^2 \, = \, \lambda \mu^2. \label{eq:mX}
\end{align}
\end{subequations}
The right-handed neutrino fermion mass is equal to the sneutrino mass $m_X$ (up to negligible soft SUSY breaking terms) because the global minimum of the inflaton potential does not break SUSY.\footnote{We neglect soft SUSY breaking terms because these are expected to be much smaller than the energy scales relevant for reheating.}

We can invert eqs.~\eqref{eq:mPhi} and \eqref{eq:mX} to express $\mu$ and $\lambda$ in terms of $m_\phi$ and $m_X$:
\begin{subequations}
\begin{align}
 \mu \, &= \, \left( \frac{ \Ne^3 }{ 6 \pi^2 \As } \right)^{\frac12} m_\phi \, \simeq \, \left( 1.3 \times 10^6 \right) m_\phi, \label{eq:mu}\\
 \lambda \, &= \, \left( \frac{ 6 \pi^2 \As }{ \Ne^3 } \right) \frac{m_X}{m_\phi^2} \, \simeq \, \left(6 \times 10^{-13}\right)\frac{m_X}{m_\phi^2} . \label{eq:lambda}
\end{align}
\end{subequations}
We can use eqs.~\eqref{eq:mu} and\eqref{eq:lambda} to express all results in terms of the physical masses $m_X$ and $m_\phi$.

\subsection{Reheat temperature}
Assuming that $\Gamma_\phi < \Gamma_\chi$ as discussed above, an approximate analytical solution of the Boltzmann equations leads to the reheat temperature (see e.g.\ \cite{valerieLeptogenesis})
\begin{align}
 T_R \, &= \, \left( \frac{36}{g_* \pi^2} \mpl^2 \Gamma_\phi^2 \right)^{\frac{1}{4}} \,
 = \, g_*^{-\frac{1}{4}} \left(  \frac{3\lambda \mpl}{4\pi^2} m_\phi m_X  \right)^{\frac{1}{2}}  \left(  1 + 12 \frac{m_X^2}{m_\phi^2}  \right)^{\frac{1}{2}} \left(  1 - 4 \frac{m_X^2}{m_\phi^2}  \right)^{^{\frac{1}{4}}},
\end{align}
with $g_* = 915/4$ in the MSSM.

\begin{figure}[ptb]
  \centering
$\begin{array}{cc}
\includegraphics[width=0.48\textwidth]{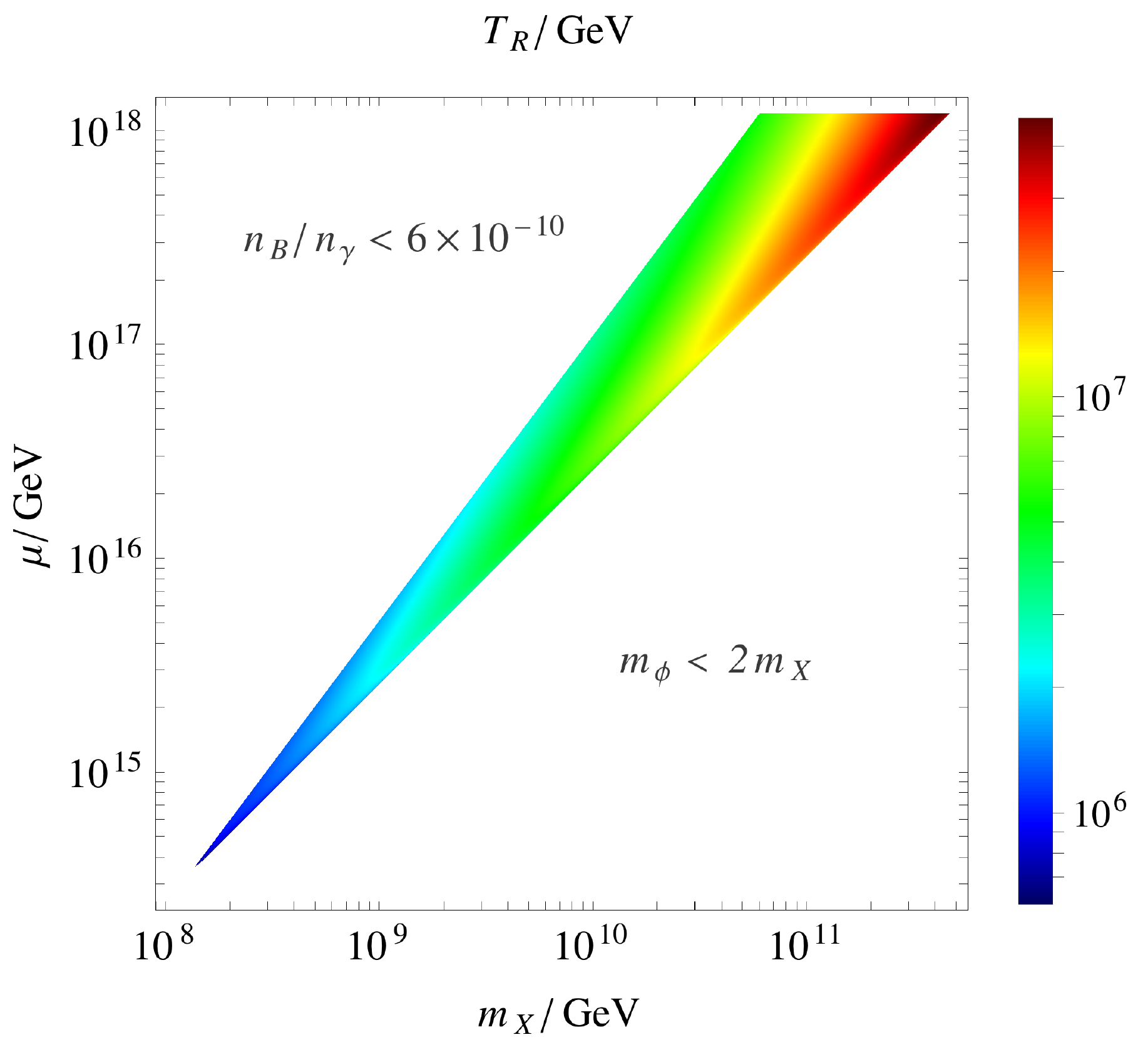} &
\includegraphics[width=0.48\textwidth]{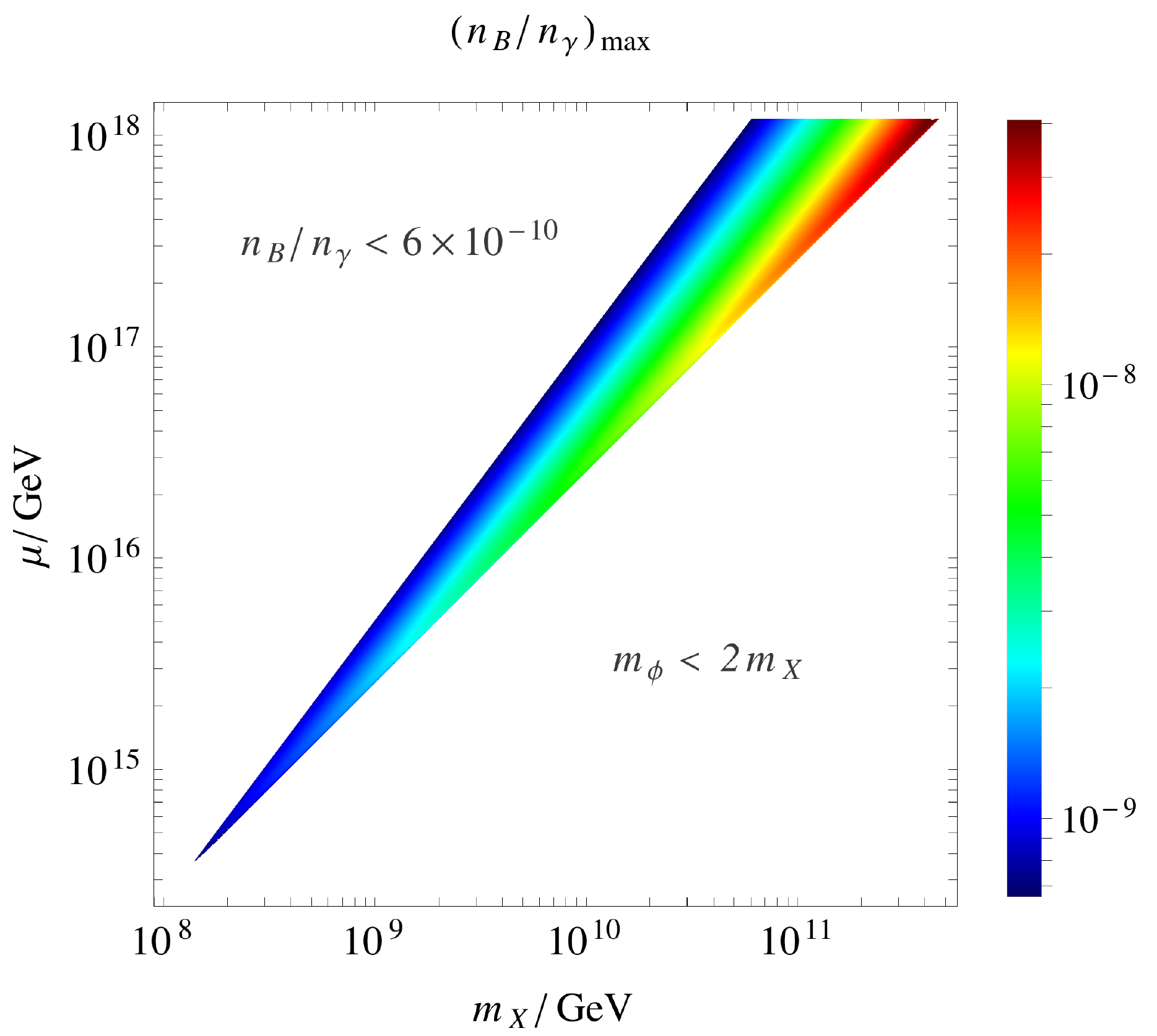}
\end{array}$
  \caption{Reheat temperature $T_R$ and upper bound on the baryon asymmetry $\frac{n_B}{n_\gamma}$ as a function of the right-handed neutrino mass $m_X$ and and inflaton vacuum expectation value $\mu$. We see that a sufficiently large baryon asymmetry can be generated for $m_X > 10^8 \, \text{GeV}$ if $\mu$ is roughly $(10^7 \times m_X)$.}
  \label{fig:TRHnB}
\end{figure}
\begin{figure}[ptb]
  \centering
$\begin{array}{cc}
\includegraphics[width=0.48\textwidth]{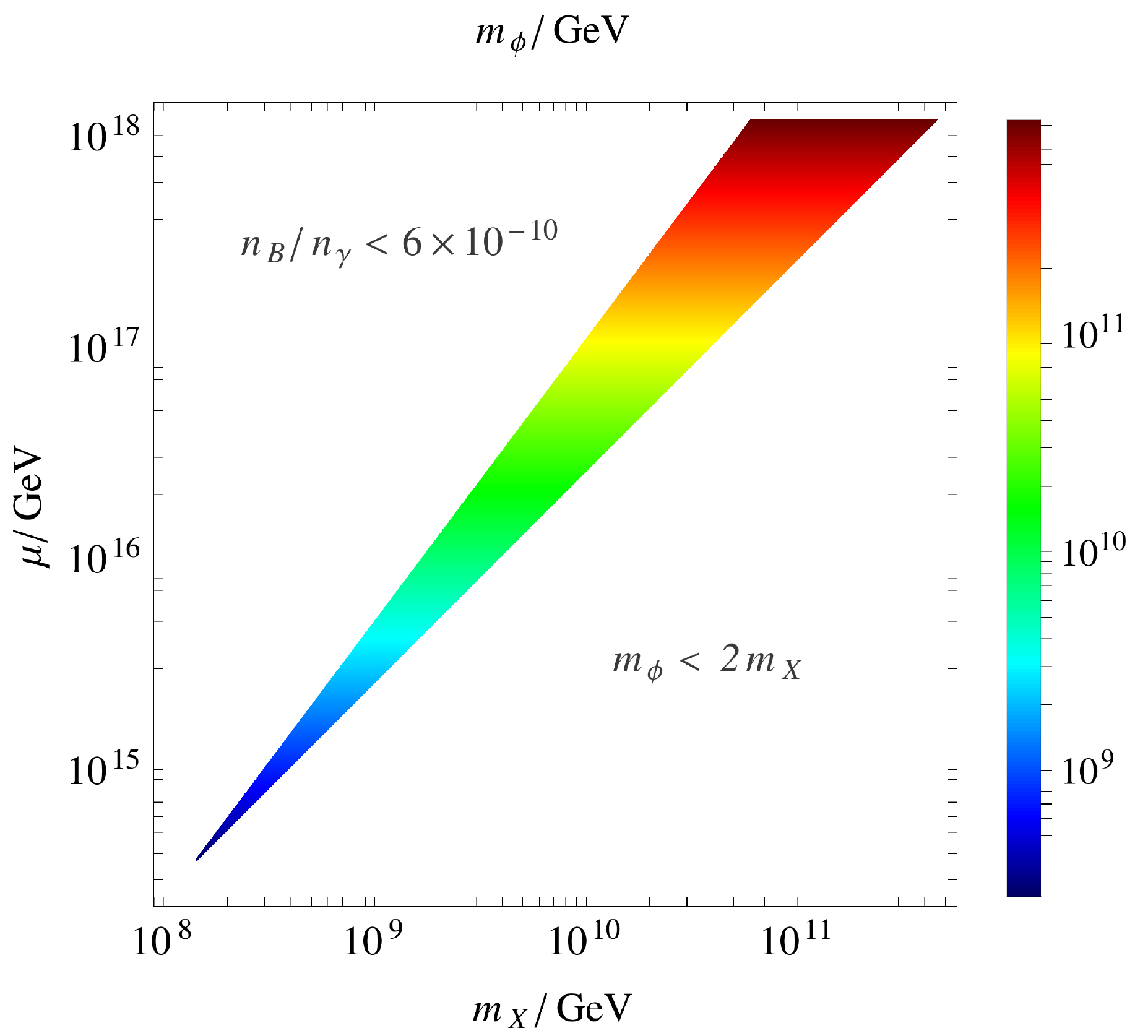} &
\includegraphics[width=0.48\textwidth]{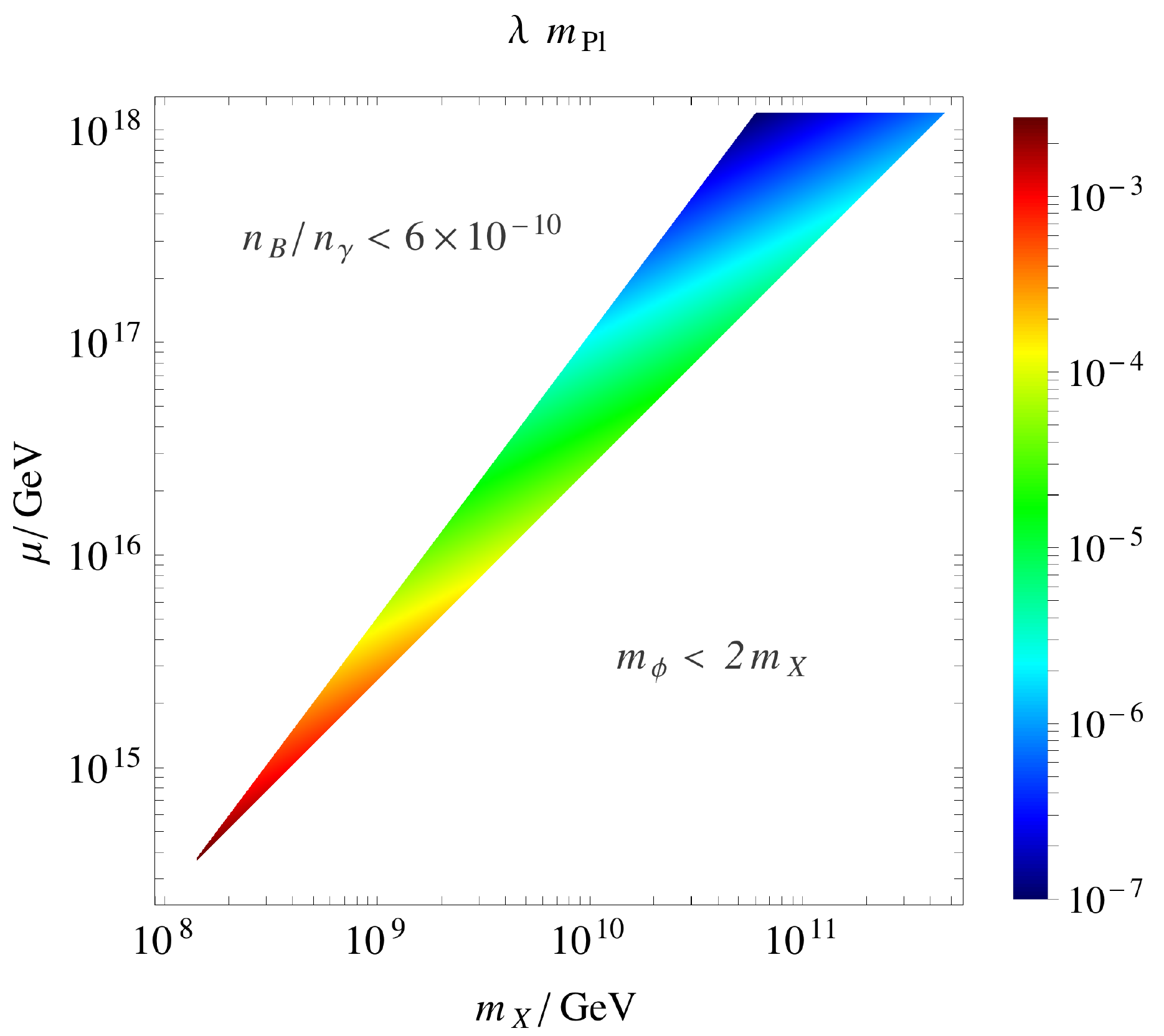}
\end{array}$
  \caption{Inflaton mass $m_\phi$ after inflation and superpotential parameter $\lambda$ as functions of the right-handed neutrino mass $m_X$ and inflaton vacuum expectation value $\mu$.}
  \label{fig:kappaLambda}
\end{figure}

For a rough estimate, we can set $\left(  1 + 12 m_X^2 / m_\phi^2  \right)^{1/2} \left(  1 - 4 m_X^2 / m_\phi^2 \right)^{1/4} \sim 1$; the actual value is between $2/3$ and $4/3$ for $m_X/m_\phi < 0.497$. With this approximation, the reheat temperature is
\begin{align}
 T_R \, \sim \, \left( \frac{9 \As \mpl}{2 \sqrt{g_*} \Ne^3} \right)^{\frac12} \frac{m_X}{\sqrt{m_\phi}} \, \simeq \, \left( \frac{ 10^{4} \, \text{GeV} }{ m_\phi } \right)^{\frac12} m_X. \label{eq:TRestimate}
\end{align}
In particular, this result implies that the right-handed (s)neutrinos are out of equilibrium during reheating ($T_R \ll m_X$) for $m_\phi \gg 10^4 \, \text{GeV}$.

The reheat temperature as a function of $m_X$ and $\mu$ is shown in fig.~\ref{fig:TRHnB} for the range of parameters for which the baryon asymmetry can be produced by nonthermal leptogenesis (see below) and $\mu < \mpl$. In this region, the reheat temperature is generally between $10^6 \, \text{GeV}$ and $10^8 \, \text{GeV}$, which can be low enough to evade the bounds on thermal gravitino production depending on the gravitino mass $m_{3/2}$ \cite{gravitinoProblem1,gravitinoProblem2,gravitinoProblem3,gravitinoProblem4,gravitinoProblem5}.

\subsection{Nonthermal leptogenesis}
The inflaton particles decay into pairs of right-handed neutrinos and sneutrinos whose out-of-equilibrium decay generates a lepton asymmetry. The generated lepton asymmetry can be estimated analogously to \cite{valerieLeptogenesis} as
\begin{align}
 \left| \frac{n_L}{s} \right| \, \simeq \, \frac{5}{2}\frac{\epsilon T_R}{m_\phi}.
\end{align}
$\epsilon$ quantifies the CP violation per sneutrino decay:
\begin{align}
 \epsilon \, \equiv \, \frac{\Gamma - \Gamma^{\text{(CP)}}}{\Gamma + \Gamma^{\text{(CP)}}},
\end{align}
where $\Gamma$ denotes the neutrino or sneutrino decay process and $\Gamma^{\text{(CP)}}$ the CP conjugate of that decay process. For hierarchical neutrinos, $\epsilon$ is bounded by \cite{cpEpsilonBound1,cpEpsilonBound2,cpEpsilonBound3}
\begin{align}
  \epsilon \, < \, \frac{3}{8\pi} \frac{\sqrt{ \Delta m_\text{atm}^2 } m_X}{\braket{h_u}^2}. \label{eq:epsilonBound}
\end{align}
This lepton asymmetry is converted into a baryon asymmetry by sphaleron processes, which introduce a conversion factor $n_B = \frac{C}{C-1}n_L$ with $C \simeq 1/3$ in the MSSM \cite{leptonToBaryon}, and $s = 7.04 \, n_\gamma$:
\begin{align}
 \left| \frac{n_B}{n_\gamma} \right| \, = \, 7.04 \left| \frac{C}{C-1} \frac{n_L}{s} \right| \, \simeq \, 8.8 \, \frac{\epsilon T_R}{m_\phi}.
\end{align}
Using the bound from eq.~\eqref{eq:epsilonBound} with $\braket{h_u} \sim 174 \, \text{GeV}$ and $\Delta m_\text{atm}^2 \simeq 3 \times 10^{-3} \, \text{eV}^2$, we find an upper bound on the baryon asymmetry (fig.~\ref{fig:TRHnB}). We can also compute a simple estimate if we approximate $T_R$ with eq.~\eqref{eq:TRestimate}. We then find
\begin{align}
 \frac{n_B}{n_\gamma} \, \lesssim \, 6\times 10^{-10} \frac{(2m_X)^2}{\sqrt{ ( 2 \times 10^{8} \, \text{GeV}) m_\phi^3 }}. \label{eq:nBnGamma}
\end{align}
If we demand that nonthermal leptogenesis generates the observed baryon asymmetry $n_B/n_\gamma = 6 \times 10^{-10}$, eq.~\eqref{eq:nBnGamma} together with $m_\phi > 2m_X$ implies a lower bound on the inflaton mass: $m_\phi > 2 \times 10^8 \, \text{GeV}$. Also, one can see in fig.~\ref{fig:TRHnB} that this condition relates the right-handed neutrino mass $m_X$ and the symmetry breaking scale $\mu$, which must be related roughly by $\mu \sim 10^7 \times m_X$.

\subsection{Consistency of $\Gamma_\phi < \Gamma_\chi$}
\label{sec:neutrinoMassScale}

For our calculations, we have assumed that $\Gamma_\phi < \Gamma_\chi$, which according to eq.~\eqref{eq:decayRateRatio} means that the Yukawa coupling must be sufficiently large. We can estimate the size of the Yukawa coupling from the seesaw formula for the light neutrino mass $m_\nu$:
\begin{align}
 m_\nu \, \sim \, \frac{ y^2 \braket{h_u}^2 }{ m_X },
\end{align}
which implies that the Yukawa coupling is
\begin{align}
y^2 \, \sim \, \left( \frac{ m_\nu }{ 1 \, \text{meV} } \right) \left( \frac{ m_X }{ 3 \times 10^{16} \, \text{GeV} } \right).
\end{align}
If we insert this in eq.~\eqref{eq:decayRateRatio} and use eq.~\eqref{eq:lambda}, we find
\begin{align}
 \frac{\Gamma_\phi}{\Gamma_\chi} \, &\sim \, \frac{\lambda m_\phi}{2 y^2}
 \, \sim \, \left( \frac{ 10^{-4} \, \text{meV} }{ m_\nu } \right) \left( \frac{ 10^8 \, \text{GeV} }{ m_\phi } \right).
\end{align}
With $m_\phi > 2 \times 10^{8} \, \text{GeV}$, the assumed decay rate hierarchy holds if the light neutrino mass generated by the seesaw mechanism is larger than about $10^{-4}$ meV, which is true for at least two of the three light neutrinos.

\section{Conclusions}

In this paper, we have proposed a new class of models of hilltop inflation where the initial conditions of the inflaton close to the hilltop are generated from ``matter field preinflation''. This is achieved via a coupling term between the inflaton and a matter field (cf. second term in eq.~\eqref{wgen}), in addition to the usual term for supersymmetric hilltop inflation. The same coupling term also opens up a decay channel for the inflaton into Standard Model fields, which allows efficient reheating of the universe. 

Our mechanism works as follows: During preinflation, the scalar component of the matter superfield (or of a D-flat combination of them) has a large vacuum expectation value which induces a mass for the inflaton of hilltop inflation, driving its vacuum expectation value to zero. The initial conditions for hilltop inflation are then generated when the preinflaton approaches a ``critical point'' where its stabilising effect ends and the fields enter a ``diffusion region'' in which quantum fluctuations have to be included. 

We have calculated and discussed the resulting initial conditions  in section~\ref{sec:ics} and found that there are two cases:  When the mass of the preinflaton lies above a certain threshold, initial conditions are generated such that the inflationary dynamics resembles almost entirely single field hilltop inflation. Single field hilltop inflation acts as an attractor solution where the initial value of the preinflaton becomes negligible when hilltop inflation starts.

The second case happens when the mass of the preinflaton is below the above-mentioned threshold. Then, the motion of the preinflaton during hilltop inflation can no longer be neglected and inflation has to be analysed as a true two field inflation model, using e.g.\ the $\delta N$ formalism. We discussed this in section~\ref{sec:numericalResults} based on a numerical analysis, where we also highlighted the differences between these two cases.

As mentioned above, in addition to providing the right initial conditions for hilltop inflation, ``matter field preinflation'' has the attractive additional feature of opening up a decay channel for the inflaton into SM fields which can lead to an efficient reheating of the universe (with high enough but not too high reheat temperature). This can help to avoid unwanted decays of the inflaton into hidden sector fields and may even allow to explain the baryon asymmetry of the universe via the non-thermal leptogenesis mechanism. 

As an example, in section~\ref{sec:reheating} we calculated reheating and non-thermal leptogenesis for an example model where one of the right-handed sneutrinos acts as the preinflaton. We found that successful non-thermal leptogenesis is indeed possible and imposes a relation between the symmetry breaking scale and the mass of the right-handed (s)neutrino after inflation, as well as lower bounds on both quantities. In 
an explicit model, these would be interesting additional constraints on the model parameters.

We conclude that the proposed class of models of ``hilltop inflation with matter field preinflation'' is a promising framework for embedding inflation into particle physics theories. 

\section*{Appendices}
\appendix

\section{Scalar potential from supergravity}
\label{appendix:scalarPotential}

In this appendix, we discuss how the scalar potential \eqref{vinf} can be obtained from the superpotential $W$ and the \kahler potential $K$.

The supergravity scalar potential for chiral superfields can be calculated from $W$ and $K$ by evaluating the F-term potential\footnote{If $\hat{\Phi}$ or $\hat{X}$ are identified with contractions of fields which are charged under a gauge symmetry, one could also have a D-term potential. In this case, inflation is assumed to happen along a D-flat (i.e.\ gauge invariant) direction in field space, which minimizes the D-term potential.}
\begin{equation}
\label{eq:localSusyVF}
 V_F = \left. e^{K/\mpl^2}\left( D_i K^{i \overline{j}} D_j^\dagger  - \frac{3}{\mpl^2} \lvert W \rvert^2  \right) \right|_{\theta=0},
\end{equation}
where $\theta=0$ means replacing all superfields with their scalar components, $K^{i \overline{j}}$ is the matrix inverse of the \kahler metric $K_{ \overline{i}j }$, and
\begin{align}
 K_{ \overline{i}j } = \frac{ \partial^2 K }{ \partial \hat{Y}^\dagger_i \partial \hat{Y}_j },\quad
 D_i = \frac{ \partial W }{ \partial \hat{Y}_i } + \frac{W}{\mpl^2} \frac{ \partial K }{ \partial \hat{Y}_i },
\end{align}
with $\hat{Y} = \{\hat{S}, \hat{\Phi}, \hat{X}\}$. We want to evaluate this expression for the superpotential given by eq.~\eqref{wgen}. We also need to specify the \kahler potential $K$. In general, the allowed operators in the \kahler potential depend on the symmetries of the model. However, for canonically normalized fields, $K$ always contains a term $\hat{Y}_i^\dagger\hat{Y}_i$ for each chiral superfield $\hat{Y}_i$. The \kahler potential can also contain products and higher powers of these terms, so we expand $K$ in powers of the modulus squared of the superfields:\footnote{The most general \kahler potential can contain other terms which are not proportional to the modulus squared, e.g. $\left(\Phi^2 + \text{h.c.}\right)$. Most of these terms are forbidden by the $U(1)_R$ symmetry for $S$ and by the $\mathbb{Z}_p$ symmetry for $\Phi$. The remaining phase-dependent terms, e.g.\ $\left(\Phi^p + \text{h.c.}\right)/\mpl^{p-2}$, are suppressed by powers of $\Phi/\mpl$ such that their influence on the dynamics is negligible for $\Phi \ll \mpl$.}
\begin{align}
 K = \hat{S}^\dagger\hat{S} + \hat{\Phi}^\dagger\hat{\Phi} + \hat{X}^\dagger\hat{X} - \frac{\kappa_S}{\mpl^2} ( \hat{S}^\dagger\hat{S} )^2 + \frac{(1 - \alpha)}{\mpl^2} (\hat{S}^\dagger\hat{S})(\hat{X}^\dagger\hat{X}) + \frac{(1 + \beta)}{\mpl^2} (\hat{S}^\dagger\hat{S})(\hat{\Phi}^\dagger\hat{\Phi}) + ...,
\end{align}
where the $...$ denote other Planck-suppressed terms that can be safely neglected, and $\alpha$, $\beta$ and $\kappa_S$ are free parameters of the theory.

Expanding the inverse \kahler metric and the prefactor $e^{K/\mpl^2}$ in powers of the fields and dropping irrelevant higher order terms, eq.~\eqref{eq:localSusyVF} takes the form
\begin{align}
 V_F &= \left. \left( \left| \frac{ \partial W }{ \partial \hat{S} } \right|^2 \left(  1 + \frac{ 4\kappa_S}{\mpl^2} \lvert \hat{S} \rvert^2 + \frac{\alpha}{\mpl^2} \lvert \hat{X} \rvert^2 - \frac{\beta}{\mpl^2} \lvert \hat{\Phi} \rvert^2   \right) + \left| \frac{ \partial W }{ \partial \hat{\Phi} } \right|^2 + \left| \frac{ \partial W }{ \partial \hat{X} } \right|^2  \right) \right|_{\theta=0} + ... \notag\\
 &= \left|  \Lambda^2 - \frac{\Phi^p}{M^{p-2}}  \right|^2 \left(  1 + \frac{4\kappa_S}{\mpl^2} \lvert S \rvert^2 + \frac{\alpha}{\mpl^2} \lvert X \rvert^2 - \frac{\beta}{\mpl^2} \lvert \Phi \rvert^2  \right)  +  4 \lambda^2 \lvert  \Phi^2 X^4  \rvert  +  4 \lambda^2 \lvert  \Phi^4 X^2  \rvert + ...
 \end{align}
 We can rewrite this potential in terms of the real fields $\chi = \sqrt{2} \lvert X \rvert$ and $\phi = \sqrt{2} \operatorname{Re}(\Phi)$. We also drop the imaginary inflaton component $\bar{\phi} = \sqrt{2} \operatorname{Im}(\Phi)$ for simplicity.\footnote{The effect of the imaginary inflaton component is discussed in section~\ref{sec:imaginaryInflatonEffects}.} This results in the scalar potential
\begin{align}
 V_F &= \Lambda^4 \left( 1 - \frac{\phi^p}{2^{p/2}\Lambda^2 M^{p-2}} \right)^2 + \frac{\lambda^2}{2} \phi^2 \chi^2 \left( \phi^2 + \chi^2 \right) - \frac{1}{2}  \frac{\beta\Lambda^4}{\mpl^2} \phi^2 + \frac{1}{2}  \frac{\alpha\Lambda^4}{\mpl^2} \chi^2 +  \frac{4\kappa_S\Lambda^4}{\mpl^2} \lvert S \rvert^2 + ...
\end{align}
We see that $S=0$ is stabilized with a mass $m_S^2 = 4\kappa_S \Lambda^4/{\mpl^2}$. Throughout this paper we assume that $\kappa_S > \frac{1}{12}$, which means that $S$ has a mass above the Hubble scale and can be neglected during inflation. Apart from this mass term, the potential is identical to the one we gave in eq.~\eqref{vinf} with the parameters
 \begin{align}
  m_\phi^2 = \beta \Lambda^4/{\mpl^2}, \quad
  m_\chi^2 = \alpha \Lambda^4/{\mpl^2}, \quad
  \mu^p = 2^{p/2} \Lambda^2 M^{p-2}.
 \end{align}
Note that $\alpha$ and $\beta$ could be either positive or negative; in this paper, we require that $\alpha > 0$ and $\beta > 0$.
 
 \section{Decay rate calculation}
\label{appendix:decayRates}
 
In this appendix we calculate the perturbative inflaton decay for our example model in section \ref{sec:reheating}, where the matter superfield $\hat{X}$ is identified with the right-handed (s)neutrino.

In the rest frame of the decaying particle, the decay rate $\Gamma$ of a particle with mass $m_\text{i}$ into two particles of mass $m_\text{f}$ is given by
\begin{align}
 \Gamma_{\text{i} \, \rightarrow \, \text{ff}} \, = \, \frac{1}{16\pi m_\text{i}} \left|  \mathcal{M}  \right|^2 \sqrt{ 1 - \frac{4m_\text{f}^2}{m_\text{i}^2} }. \label{eq:decayRateFormula}
\end{align}
To get the total decay rate, we must sum over the decay rates for each channel. This includes summing over spin polarizations for final-state fermions and averaging over spin polarizations for initial-state fermions.

\subsection{Inflaton decay rate $\Gamma_\phi$}

\begin{figure}[btp]
  \centering
$\begin{array}{ccc}
\includegraphics[width=0.32\textwidth]{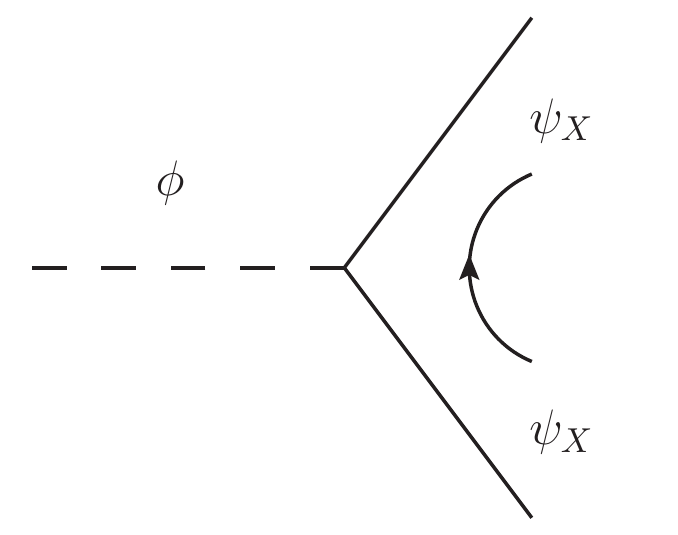} &
\includegraphics[width=0.32\textwidth]{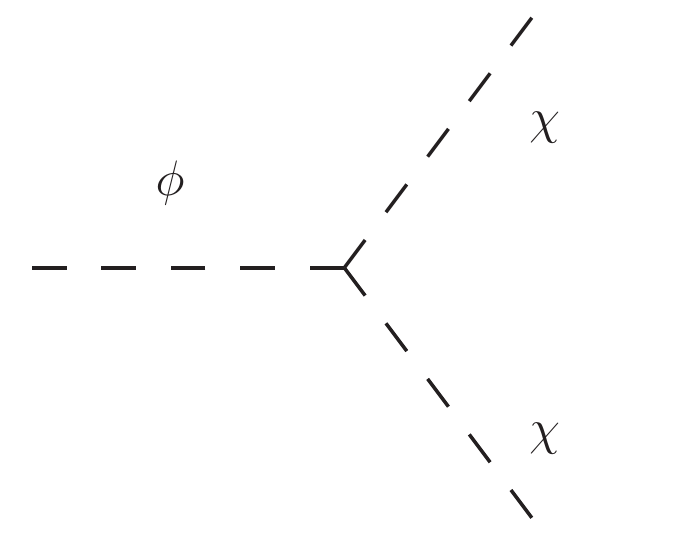} &
\includegraphics[width=0.32\textwidth]{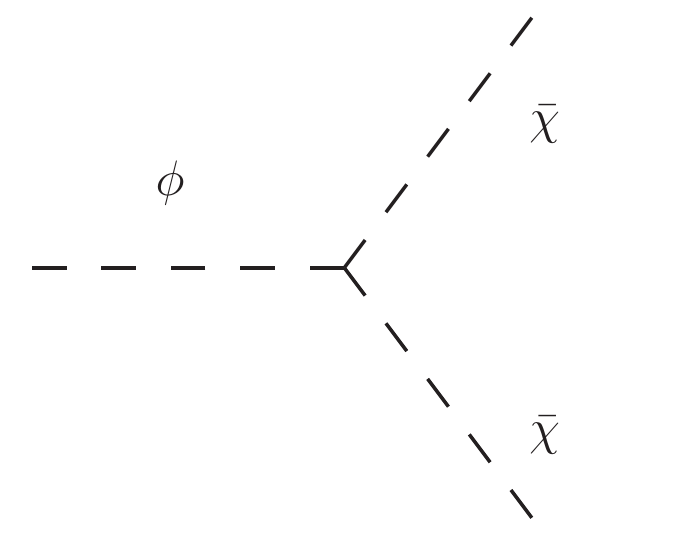}
\end{array}$
  \caption{Decay channels for the inflaton $\phi$ into right-handed neutrinos $\psi_X$ and right-handed sneutrinos $\chi$, $\chiI$.}
  \label{fig:phiDecay}
\end{figure}

For the decay of the inflaton $\phi$, there are two diagrams for the decay into scalar sneutrinos\footnote{As a complex scalar field, the sneutrino consists of two real components, which we denote by $\chi$ and $\chiI$.} and one diagram for the decay into right-handed neutrinos, see fig.~\ref{fig:phiDecay}. The matrix elements for these decays are\footnote{The Feynman rules for Majorana fermions can be found e.g.\ in \cite{majoranaFeynmanRules}.}
\begin{subequations}
\begin{align}
  i\mathcal{M}_{\phi \, \rightarrow \, \text{fermions} } \, &= \, -i \sqrt{\lambda_i m_X} \, \bar{u}(p_1,s_1) v(p_2,s_2) ,\\
  i\mathcal{M}_{\phi \, \rightarrow \, \text{scalars}} \, &= \, -4i \sqrt{ \lambda_i m_X^3 },
\end{align}
\end{subequations}
with $p_1 = (\frac{m_\phi}{2}, \vec{p_1})$ and $p_2 = (\frac{m_\phi}{2}, -\vec{p_1})$. We get the total decay rate by summing over eq.~\eqref{eq:decayRateFormula} for each of these three decay channels:
\begin{align}
 \Gamma_\phi \,
 &= \, \frac{\lambda}{8\pi} m_\phi m_X \left(  1 + 12 \frac{m_X^2}{m_\phi^2}  \right)\left(  1 - 4 \frac{m_X^2}{m_\phi^2}  \right)^{1/2}.
\end{align}

\subsection{Sneutrino decay rate $\Gamma_\chi$}

\begin{figure}[btp]
  \centering
$\begin{array}{cccc}
\includegraphics[width=0.24\textwidth]{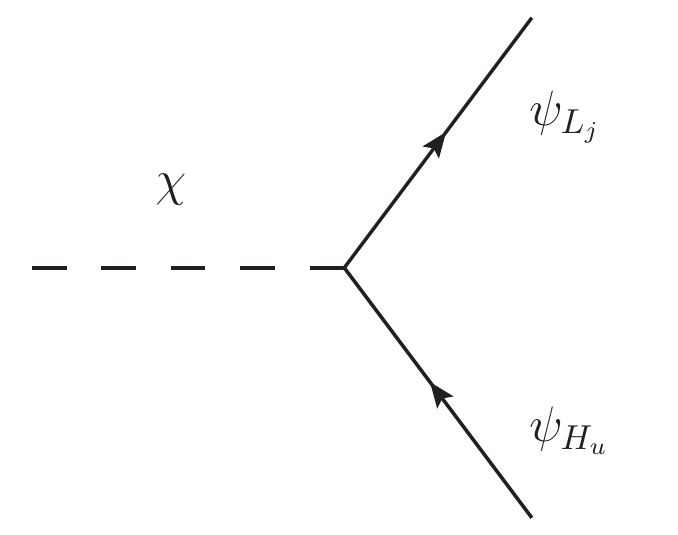} &
\includegraphics[width=0.24\textwidth]{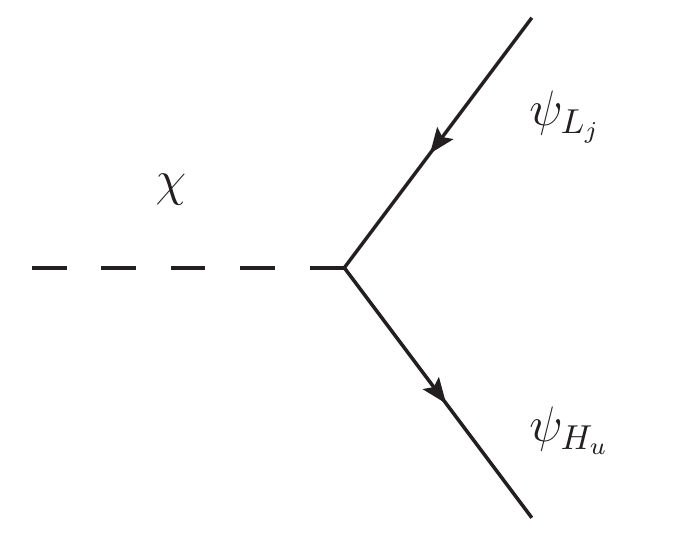} &
\includegraphics[width=0.24\textwidth]{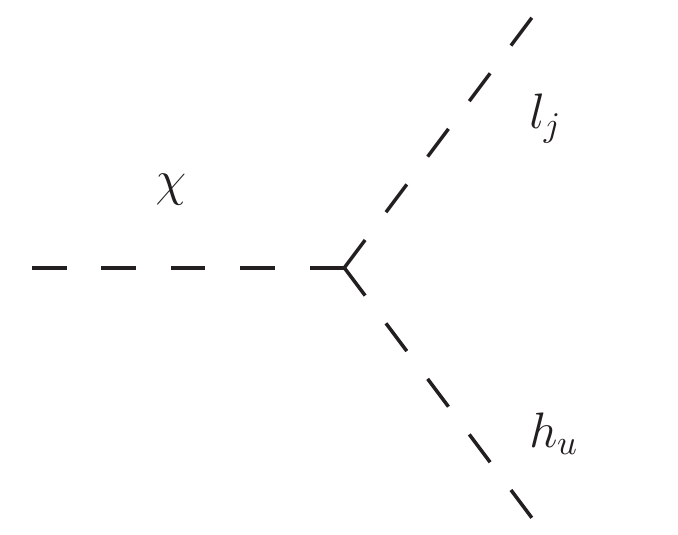} &
\includegraphics[width=0.24\textwidth]{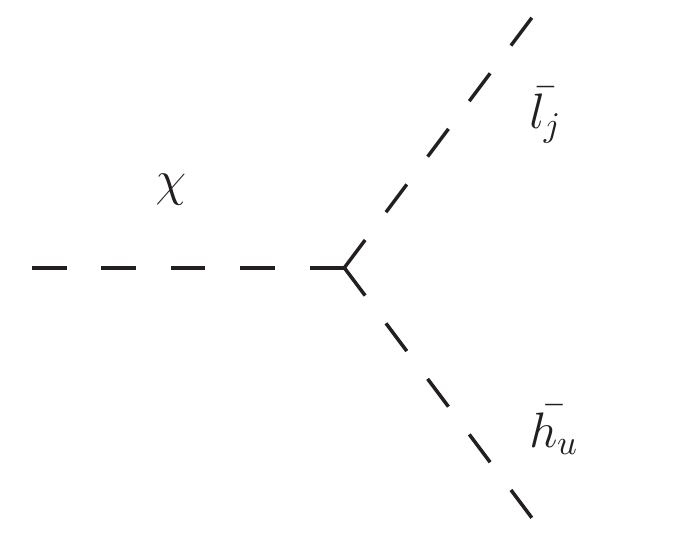} \\
\includegraphics[width=0.24\textwidth]{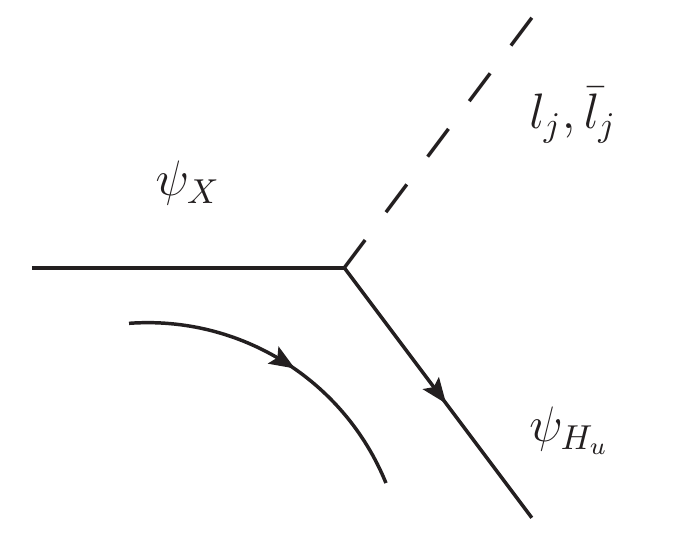} &
\includegraphics[width=0.24\textwidth]{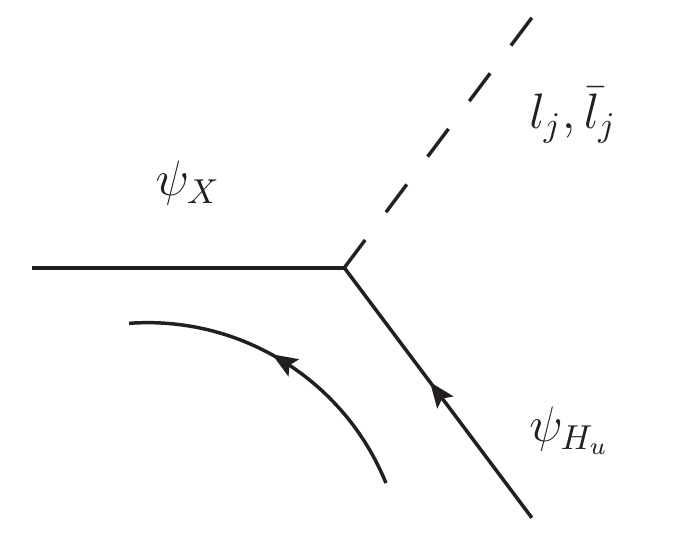} &
\includegraphics[width=0.24\textwidth]{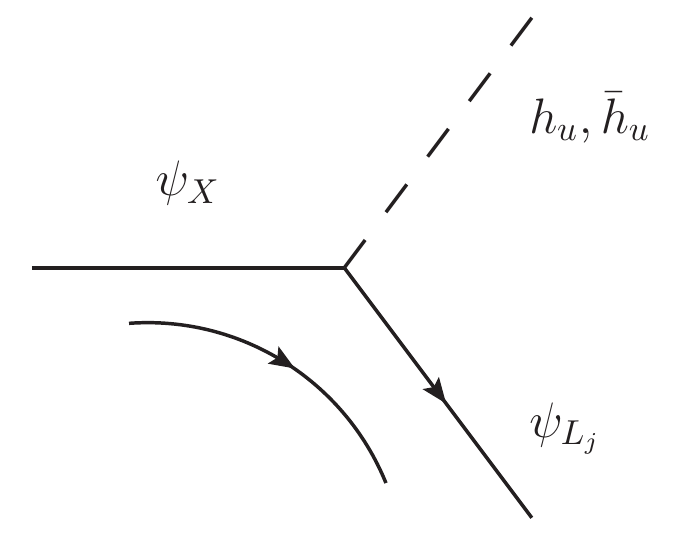} &
\includegraphics[width=0.24\textwidth]{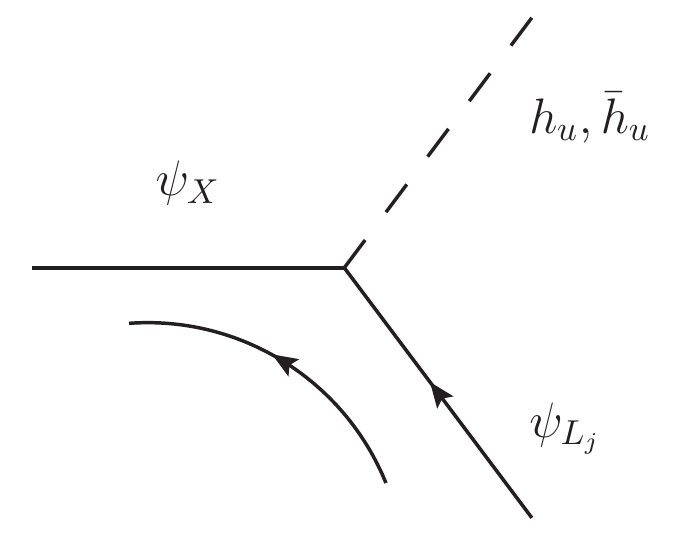}
\end{array}$
  \caption{Decay channels for the right-handed sneutrino $\chi$ and the right-handed neutrino $\psi_X$ into left-handed (s)leptons and Higgs(ino) particles.}
  \label{fig:chiDecay}
\end{figure}

For the decay of the right-handed sneutrino $\chi$, we have 12 decay channels into scalars and 12 decay channels into fermions, see fig.~\ref{fig:chiDecay}. Note that each of the diagrams in fig.~\ref{fig:chiDecay} stands for six separate processes due to the decay into two different $SU(2)$ components (index suppressed) and three generations (index $j$). The matrix elements for each of these decays are
\begin{subequations}
\begin{align}
 i\mathcal{M}_{\chi \, \rightarrow \, \text{fermions}} \, &= \, \frac{-i}{\sqrt{2}} \, y_{ji} \, \bar{u}(p_1,s_1) \frac{1 + \gamma^5}{2} v(p_2,s_2), \\
 i\mathcal{M}_{\chi \, \rightarrow \, \text{scalars}} \, &= \, \frac{-i}{\sqrt{2}} \, y_{ji} m_X,
\end{align}
\end{subequations}
with $p_1 = (\frac{m_X}{2}, \vec{p_1})$ and $p_2 = (\frac{m_X}{2}, -\vec{p_1})$. We get the total decay rate by summing over eq.~\eqref{eq:decayRateFormula} for each of these eight decay channels:
\begin{align}
  \Gamma_\chi \, &= \, \frac{ \sum_j \lvert y_{ji} \rvert^2 }{4\pi} m_X,
\end{align}
where we neglected $m_\text{f} \sim m_{H_u} \sim m_{L_j} \ll m_X$. One can easily verify that the other scalar sneutrino component $\chiI$ has the same decay rate.

\subsection{Neutrino decay rate $\Gamma_{\psi_X}$}

The right-handed neutrino $\psi_X$ can decay into left-handed (s)leptons and Higgs(ino) particles through the diagrams in fig.~\ref{fig:chiDecay}. The four diagrams correspond to 16 different processes for each $j$: one factor of two is due to the $SU(2)$ contraction and another factor of two is because the final state can contain each of the two real components of $L_j$ or $H_u$. The matrix elements are
\begin{align}
 \left| \mathcal{M}_{\psi_X \, \rightarrow \, \text{scalar}+\text{fermion}} \right| \, &= \, \frac{1}{\sqrt{8}} \left| y_{ji} \, \bar{u}(p_\text{f},s_\text{f}) \left( 1 + \gamma^5 \right) u(p_\text{i},s_\text{i}) \right|
\end{align}
for the decay into particles, with $p_\text{f} = (\frac{m_X}{2},\vec{p_\text{f}})$ and $p_\text{i} = (m_X, \vec{0})$. The decay into antiparticles is nearly the same, but the $u$-spinors are replaced by $v$-spinors and $(1+\gamma^5)$ is replaced by $(1-\gamma^5)$, which still leads to the same squared matrix element.

The total decay rate is the sum over the decay rates for all processes. We must also include a factor $\frac{1}{2}$ for averaging over incoming spin polarizations. The result is
\begin{align}
 \Gamma_{\psi_X} \, &= \, \frac{\sum_j \left| y_{ji} \right|^2}{4\pi} m_X,
\end{align}
which is identical to the decay rate $\Gamma_\chi = \Gamma_{\chiI}$ of the two sneutrino components.

\end{document}